\documentclass[
        12pt,
        notitlepage, 
        tightenlines,
        eqsecnum,
        floats,
        floatfix,
        nofootinbib,
        amsmath,amssymb,
        aps,prd,
        longbibliography]{revtex4-2}

\usepackage[utf8]{inputenc}
\DeclareUnicodeCharacter{0302}{}
\usepackage{microtype}
\usepackage{comment}
\usepackage[normalem]{mathtools,ulem}
\usepackage{graphicx}
\usepackage{subcaption}
\usepackage{physics}
\usepackage{float}
\usepackage{amssymb}
\usepackage[align,text]{tensor}
\usepackage{dirtytalk}
\usepackage[usenames,dvipsnames,x11names,svgnames]{xcolor}
\usepackage{booktabs,multirow}
\usepackage{mathrsfs}
\usepackage{amsthm}

\newcommand\myshade{85}
\colorlet{mylinkcolor}{Red1}
\colorlet{mycitecolor}{Blue1}
\colorlet{myurlcolor}{Blue1}

\usepackage{hyperref}
\hypersetup{
unicode,
breaklinks, 
colorlinks = true, 
linkcolor  = mylinkcolor!\myshade!black,
citecolor  = mycitecolor!\myshade!black,
urlcolor   = myurlcolor!\myshade!black, 
}

\usepackage{cleveref}
\usepackage{bm,nicefrac,xfrac,dsfont}
\usepackage{bbm}

\DeclareMathOperator{\sgn}{sgn}

\begin{document}

\title{A loop quantization of the marginally bound Lemaître-Tolman-Bondi dust model}

\author{Luca Cafaro}
\email{lcafaro@fuw.edu.pl}

\author{Farshid Soltani}
\email{f.soltani@uw.edu.pl}

\address{Faculty of Physics, University of Warsaw, Pasteura 5, 02-093 Warsaw, Poland}


\begin{abstract}
\vspace*{.3cm}

We present a loop quantization of the marginally bound Lemaître-Tolman-Bondi (LTB) model, describing the gravitational collapse of pressureless dust in spherical symmetry. The full quantum LTB model is constructed as a collection of non-interacting shells, each governed by an individual single-shell loop quantum dynamics. We show that the single-shell evolution is non-singular and that wave packets initially peaked on a collapsing trajectory undergo a bounce at Planckian energy densities and subsequently follow an expanding classical trajectory, resolving the classical central curvature singularity. We also compare the loop quantum theory with the Wheeler-DeWitt quantization of the same model. Finally, we briefly comment on the accuracy of the loop quantum gravity effective theory in reproducing this full quantum dynamics.
Specifically, we find that initially collapsing wave packets generically develop an interference pattern at the bounce, which suppresses the accuracy of the effective theory near the center of the dust cloud.

\end{abstract}

\maketitle

\section{Introduction} 
\label{sec:1}

Due to the non-linear nature of general relativity (GR), analytic solutions typically rely on the presence of symmetries. In the context of gravitational collapse, spherical symmetry is commonly imposed, as it provides a simple yet non-trivial framework for the analysis. The first landmark result in this setting was obtained by Oppenheimer and Snyder in their seminal work~\cite{OS}, where they studied the collapse of a homogeneous pressureless dust cloud and demonstrated its inevitable culmination in black hole formation. Given its far-reaching astrophysical implications, this model rapidly attracted considerable attention and it was soon extended to describe inhomogeneous dust profiles, giving rise to the now well-established Lemaître-Tolman-Bondi (LTB) model~\cite{Lemaitre, Tolman, Bondi} (see~\cite{Plebanski_Krasinski_2006} for a modern review). Building on these results, increasingly general and physically realistic scenarios were explored, including models with different matter sources~\cite{Vaidya, Holm_Kupershmidt}, non-vanishing pressure~\cite{Lasky_Lun}, anisotropic stresses~\cite{cadogan2024}, heat exchange~\cite{Govender}, and beyond. Yet regardless of the complexity of the fluid under consideration, classical general relativity invariably dictates that gravitational collapse ends in a curvature singularity.

It is widely believed that the quantum nature of gravity would provide the key to singularity resolution. In the absence of a fully developed quantum gravity theory, a common strategy to gain insight into the microscopic structure of gravity is to study suitably modified versions of GR. Such theories, which we broadly refer to as ``semiclassical'', are typically motivated by existing quantum gravity frameworks or by quantum mechanical principles such as the generalized uncertainty principle~\cite{Fragomeno:2024tlh, Barca:2023shv}. On one hand, semiclassical theories carry the significant advantage of preserving the classical mathematical structure, in that they yield a set of differential equations that reduce to Einstein's field equations in the appropriate classical limit. On the other hand, they merely mimic an expected quantum behavior without maintaining an explicit connection to the quantum theory from which they draw inspiration. In many, though not all, cases these theories successfully remove the Schwarzschild singularity. Loop quantum gravity (LQG) effective models, for instance, predict a bounce at Planckian scales~\cite{Bambi, Rovelli:2014cta, BenAchour:2020mgu, Kelly_2020_2, Giesel_2021, Bobula:2023kbo, Lewandowski, Han:2023wxg, Fazzini:2023scu, GieselEmbedding, Bobula:2024ywp}. However, they do not necessarily lead to a completely non-singular collapse, as shell-crossing singularities are found to persist~\cite{Husain:2021ojz,Husain_2022,Fazzini_2024,Cipriani:2024nhx,Bobula:2024chr,GieselGeneralized,Giesel:2024mps,Cafaro,Giesel:2026pjj, bobula2026}. A conceptually more satisfactory, albeit technically more demanding, approach to probing the quantum physics of gravitational collapse consists in directly quantizing the symmetry-reduced models. The main obstacle to this program is that the inherent complexity of these quantum theories hinders detailed dynamical investigations even in the simplest cases.

In this article, pursuing the latter approach, we introduce a (loop) quantum theory for the marginally bound LTB model. A well-known difficulty in the quantization of non-isotropic models is the presence of radial derivatives in the Hamiltonian, which are notoriously difficult to handle at the quantum level. Indeed, while the kinematical framework of a quantum LTB model can be established by following the frameworks developed in~\cite{Bojowald_2004,Bojowald:2005cb,Gambini_2014,Gambini:2020nsf} and~\cite{Husain:2021ojz,Husain_2022}, its dynamical content proves largely intractable. To overcome this obstacle, we first develop the quantum theory of each individual LTB matter shell, and then construct a quantum theory for the full LTB model starting from the single-shell theory, under the assumption that the classical decoupling between shells is preserved at the quantum level.

The Hamiltonian theory of a single LTB matter shell can be conveniently obtained by imposing the so-called LTB condition~\cite{Bojowald_2008,Giesel_2021,GieselEmbedding} on the Hamiltonian theory of the marginally bound LTB model. This condition enters as an additional first-class constraint that reduces the degrees of freedom of the full model to those of a single shell. Upon complete gauge fixing, the resulting physical Hamiltonian of the single-shell model takes the same form as the Hamiltonian constraint of a spatially flat Friedmann universe filled with dust, thereby allowing the quantum theory to be developed using the techniques of loop quantum cosmology (LQC)~\cite{Ashtekar:2011ni}. Throughout this work, we will refer to this quantization procedure, applied in the present symmetry-reduced context, as ``loop quantization''.

The same single-shell theory was derived in~\cite{Kiefer_2019} from a Lagrangian perspective and subsequently quantized via the Wheeler-DeWitt (WDW) approach. In the present work, we reformulate this theory in terms of phase space variables adapted to the loop quantization, and compare its physical predictions with those of the loop quantum theory. In addition, we assess the accuracy with which the LQG semiclassical theory reproduces the dynamics of the full quantum theory.

The classical theory is reviewed in \cref{sec:2}, where the reduction to the single-shell model and the gauge fixing procedure are discussed in \cref{sec:2.1}. The quantum theory is then developed in \cref{sec:3}, beginning with the Wheeler-DeWitt quantization of the single-shell model in \cref{sec:3.1}, followed by its loop quantization in \cref{sec:3.2}, and culminating in the loop quantum theory of the full marginally bound LTB model in \cref{sec:3.3}. The paper closes with \cref{sec:4}, where the main results are discussed and the limitations of the model are addressed.

\section{Classical theory} \label{sec:2}

We want to study the gravitational collapse of a ball of dust in spherical symmetry, known as the LTB model. As rotation necessarily vanishes for a perfect fluid acting as the source of Einstein's field equations in spherical symmetry~\cite{Plebanski_Krasinski_2006}, we can consider the following action $S$ for our system
\begin{equation}
    S = S_{\mathrm g} + S_{\mathrm{dust}} = \frac{1}{16\pi G}\int_{\mathcal{M}} d^4 x \sqrt{-g}\, R -\frac{1}{2} \int_{\mathcal{M}} d^4 x \sqrt{-g}\, \rho \, \Big(g^{\mu\nu} \partial_\mu T \partial_\nu T + 1\Big) ,
\end{equation}
where $S_{\mathrm g}$ is the Einstein-Hilbert action for the gravitational field $g_{\mu\nu}$, $R$ being the Ricci scalar associated to it, and $S_{\mathrm{dust}}$ is the action for irrotational dust discussed for example in~\cite{BrownKuchar,Husain_2012}. From the definition of the stress-energy tensor, and using the equations of motion (EOMs) for the fields $\rho$ and $T$, we obtain
\begin{equation}
    T_{\mu\nu}^{\mathrm{dust}} = \rho\, \partial_\mu T \partial_\nu T
\end{equation}
showing that $\rho$ represents the energy density of the dust field and $U_\mu \coloneq -\partial_\mu T$ its 4-velocity field, with the negative sign ensuring that $U^\mu$ is future oriented. Furthermore, from the equation of motion for $\rho$ we also have $U^\mu T_{,\mu} = 1$, telling us that $U^\mu$ is correctly normalized and the field $T$ represents the proper time along the flow lines of the dust.

The first step towards a Hamiltonian description of the system is the ADM decomposition of the action. Assuming spacetime to be globally hyperbolic, which is necessary to have a well-posed initial value problem, and it implies that $\mathcal{M}$ is diffeomorphic to $\mathbb{R} \times \Sigma$ for a three-manifold $\Sigma$ of arbitrary topology, we can foliate spacetime into spacelike hypersurfaces by 
\begin{equation}
    X:\mathbb{R} \times \Sigma \rightarrow \mathcal{M} , \quad\quad\quad\quad (t,x) \mapsto X(t,x),
\end{equation}
where $t$ is a time function whose level surfaces give the spatial leaves of the foliation and $x^a$, $a=1,2,3$, are coordinates on these spatial leaves. We can then define a basis $\{\partial_t,\partial_{(a)}\}$ adapted to the foliation, with $\partial_t^\mu \coloneq \dot{X}^\mu$, the dot representing differentiation with respect to $t$, and $\partial_{(a)}^\mu \coloneq X\indices{^\mu_{,a}}$. 
The decomposition of $\partial_t^\mu$, also known as the deformation vector, in terms of the normal $n^\mu$ to the spatial leaves and the basis $\partial_{(a)}^\mu$ on it, namely
\begin{equation} \label{ADM_lapse_shft}
    \partial_t^\mu = N n^\mu + N^a \partial_{(a)}^\mu,
\end{equation}
defines the lapse function $N(t,x)$ and the shift vector $N^a(t,x)$. We assume the normal $n^\mu$ to be future oriented with respect to the coordinate $t$, which constrains the lapse to be strictly positive $N>0$. In these coordinates, the line element takes the ADM form
\begin{equation} 
    ds^2 = -N^2 \, dt^2 + q_{ab} \,\left(dx^a +N^a dt\right) \left(dx^b +N^b dt\right), 
\end{equation}
where $q_{ab}(t,x)$ is the 3-metric on the spatial leaves.

If spacetime is also spherically symmetric, as we will assume from now on, then we can always choose a foliation that is adapted to this symmetry. Namely, we can always choose spatial leaves containing entire orbits of the symmetry group and use coordinates on these orbits as spatial coordinates. This means having $x^a=(x,\theta,\varphi)$, where $x$ is an arbitrary radial coordinate and $(\theta,\varphi)$ are the angular coordinates adapted to the spherical symmetry. As a consequence, the only non-vanishing component of the shift vector $N^a$ is the radial component $N^x$, and the line element takes the simplified form
\begin{equation} \label{metric}
    ds^2 = -N^2 \, dt^2 +f^2 \left(dx +N^x \, dt\right)^2 + g^2 d\Omega^2  , 
\end{equation}
where $d\Omega^2$ is the line element of the unit 2-sphere and $f^2(t,x)$ and $g^2(t,x)$ are the only two independent components of the 3-metric $q_{ab}(t,x)$.

It is then straightforward to rewrite the total action $S$ in metric ADM form, that is, in terms of the foliation-adapted coordinates $(t,x,\theta,\varphi)$ and the metric in \cref{metric}. However, as we will be interested in a loop quantization of the model, we need to reformulate the theory in terms of triad fields $e^i_a(t,x)$ and their inverses $e^a_i(t,x)$, known as cotriads, defined in such a way that $q_{a b}=e^i_a e^j_b \delta_{ij}$ is satisfied. In spherical symmetry, for the metric in \cref{metric}, we have~\cite{Kelly_2020_1}
\begin{equation} \label{cotriad}
    e_a^i \tau_i dx^{(a)} =  f \, \tau_1 \,dx + g \, \tau_2 \,d\theta + g\, \sin{\theta} \, \tau_3 \,d\varphi,
\end{equation}
where $\tau_i\coloneq-\frac{i}{2}\sigma_i$, $\sigma_i$ being the Pauli matrices, is a basis for $\mathfrak{su}(2)$. Notice that this expression is not the most general form allowed by spherical symmetry. In fact, it does not contain any new (gauge) degree of freedom associated with the internal $\mathrm{SU}(2)$ gauge symmetry of the triad. In spherical symmetry, the $\mathrm{SU}(2)$ internal symmetry is broken down to a $\mathrm{U}(1)$ internal symmetry, and the triad should have three independent components: Two physical degrees of freedom ($f$ and $g$) and one gauge component. The most general form of the triad allowed by spherical symmetry is given, for example, in~\cite{Bojowald_2004}. As shown in~\cite{Gambini_2014}, the expression in \cref{cotriad} is obtained out of the general form by gauge-fixing the residual $\mathrm{U}(1)$ internal symmetry. As we will be interested in gauge-fixing as much symmetry as possible, we can construct our theory starting from \cref{cotriad} without any loss of generality.

The variables used in loop quantization are not actually the triads, but the densitized triads $E^a_i\coloneq\sqrt{q}\,e^a_i$. Explicitly, one has
\begin{equation}
    E^a_i \tau^i \partial_{(a)}=E^x \sin{\theta} \,\tau_1 \, \partial_x + E^\varphi \sin{\theta} \,\tau_2 \, \partial_\theta + E^\varphi \tau_3 \, \partial_\varphi \, ,
\end{equation}
where we defined $E^x\coloneq\sgn\!\big(f\big) \,g^2$ and $E^\varphi\coloneq\abs{f} g$. Notice that, as usual, while the metric is insensitive to the signs of $f$ and $g$, the (densitized) triad is not. In the following, as we will account for possible flips of the triad orientation, $f$ and $g$ can be either positive or negative functions. The line element expressed in these variables reads
\begin{equation} \label{metricE}
    ds^2 = -N^2 \, dt^2 + \frac{\:(E^\varphi)^2}{\abs{E^x}} \left(dx +N^x \, dt\right)^2 + \abs{E^x} \,d\Omega^2  . 
\end{equation}
Using the Holst action as gravitational action $S_{\mathrm g}$ in the tetrad formalism, the variable conjugate to the densitized triad turns out to be the Ashtekar-Barbero connection $A_a^i=\omega_a^i+\gamma K_a^i$, where
\begin{equation}
    \omega_a^i=\frac{1}{2} \epsilon\indices{^i_{j k}} e^{b k} \left( \partial_a e^j_b-\partial_b e_a^j + e^{c j}e_{a m} \partial_c e_b^m \right) ,
\end{equation}
is the spin connection, $\gamma \in \mathbbm{R}$ is the Barbero-Immirzi parameter, and $K_a^i=K_{a b}e^{b i}$, $K_{a b}$ being the extrinsic curvature of the spatial leaves of the foliation. Putting all this together, we have
\begin{align} \label{connection}
    \begin{split}
        A^i_a \tau_i dx^{(a)} =& \, \gamma K_x^1 \tau_1 dx + \left(\frac{\partial_x E^x}{2 E^\varphi} \tau_3+\gamma K_\theta^2 \tau_2\right) d\theta  \\
       & +\left(\cos{\theta} \tau_1 - \frac{\partial_x E^x}{2 E^\varphi}  \sin{\theta} \tau_2  + \gamma K^3_\varphi  \tau_3 \right) d\varphi
    \end{split}\\
    \begin{split}
        =& \, K_x \tau_1 dx + \left( \frac{\partial_x E^x}{2 E^\varphi} \tau_3+ K_\varphi \tau_2\right) d\theta  \\
       &+\left(\cot{\theta} \tau_1 - \frac{\partial_x E^x}{2 E^\varphi}  \tau_2 +  K_\varphi \tau_3 \right)\sin{\theta} \,d\varphi \, ,
    \end{split} 
\end{align}
where 
\begin{align}
    K_x\coloneq\gamma K_x^1=&\frac{\gamma}{N} \left(-N^x \partial_xf-f \partial_x N^x + \dot{f}\right),\\
    K_\varphi\coloneq\gamma K_\theta^2 =&\frac{\gamma}{N}\left(-N^x \partial_x g + \dot{g}\right),
\end{align}
and $\gamma K_\varphi^3 =  K_\varphi \sin{\theta} $. Importantly, the products $K_x E^x$ and $K_\varphi E^\varphi$ are independent of the orientation of the triad (i.e., the signs of $f$ and $g$). The canonical form of the gravitational action $S_{\mathrm g}$ is then given by
\begin{equation} \label{actiong}
    S_{\mathrm g} =\int dt \int dx \left[ \frac{\dot{K}_x E^x + 2 \dot{K}_{\varphi} E^{\varphi}}{2 \gamma G} - 
          N \mathcal{H}^{(\mathrm{g})}  -N^x \mathcal{H}_x^{(\mathrm{g})} \right]  ,
\end{equation}
with
\begin{subequations} \label{constraintsg}
    \begin{align}
        \mathcal{H}^{(\mathrm{g})}
            =&-\frac{1}{2 \gamma^2 G} \left[ 2 K_x K_{\varphi} \frac{E^x E^\varphi}{\sqrt{\abs{E^x}} \abs{E^\varphi}}+\frac{\abs{E^{\varphi}}}{\sqrt{\abs{E^x}}} \left(K_{\varphi}^2+\gamma^2\right)\right] \label{H_g} \\ \notag 
            &+\frac{1}{8 G} \frac{\left(\partial_x E^x \right)^2}{\abs{E^\varphi}\sqrt{\abs{E^x}}}+\frac{1}{2 G}  \frac{E^x E^\varphi}{\sqrt{\abs{E^x}} \abs{E^\varphi}} \partial_x \left(\frac{\partial_x E^x}{E^\varphi}\right) ,  \\
        \mathcal{H}_x^{(\mathrm{g})} &= \frac{1}{2 \gamma G}\left(2 E^\varphi \partial_x K_{\varphi} - K_x \partial_x E^x\right). \label{Hx_g} 
    \end{align}
\end{subequations}
%
%
%
%
%
%
Using the relation 
\begin{equation}
    g^{\mu\nu} = - n^\mu n^\nu + q^{ab} \partial^\mu_{(a)} \partial^\nu_{(b)}
\end{equation}
coming out of the ADM decomposition, it is straightforward to rewrite the dust action as
\begin{equation}
    S_{\mathrm{dust}} = -\frac{1}{2} \int dt \int d^3x \, N \sqrt{q}\, \rho \, \Big(- U^2_n + q^{ab} U_a U_b + 1\Big)
\end{equation}
where $U_a\coloneq \partial^\mu_{(a)} U_\mu$ and
\begin{equation}
    U_n \coloneq n^\mu U_\mu = - \frac{1}{N} \big( \dot{T} - N^a \partial_a T \big)=- \frac{1}{N} \big( \dot{T} - N^x \partial_x T \big)\,.
\end{equation}
We can then obtain the momenta $P_T$ and $P_\rho$ canonically conjugate to our dust fields $T$ and $\rho$ as
\begin{equation} \label{pT}
    P_T \coloneq \frac{\delta S_{\mathrm{dust}}}{\delta \dot{T}} = -  \sqrt{q}\, \rho \, U_n\,,
\end{equation}
\begin{equation} \label{prho}
    P_\rho \coloneq \frac{\delta S_{\mathrm{dust}}}{\delta \dot{\rho}} = 0\,.
\end{equation}
The angular dependence of $P_T$ can be conveniently factored out by writing $P_T = p_T \sin \theta$. Importantly, as both $U^\mu$ and $n^\mu$ are future-directed timelike vectors, $U_n$ is necessarily negative, and it follows that $\sgn(p_T)=\sgn(\rho)$. In our scenario, the dust field represents a physical matter field modeling a collapsing star and is therefore subject to the standard matter energy conditions. We must then require $\rho>0$, which in turn implies $p_T>0$, in agreement with~\cite{BrownKuchar}. As the dust action does not depend on $\dot{\rho}$, $P_\rho$ is constrained to vanish, as expressed by \cref{prho}, introducing a primary constraint in the Hamiltonian theory. The complete Hamiltonian analysis of the dust action, also in the presence of non-vanishing rotation, can be found in~\cite{Giesel_2010}. It can be easily shown that the conservation in time of this primary constraint produces a secondary constraint, which furthermore forms a second-class set with the primary constraint. The two constraints can then be solved for $(\rho,P_\rho)$ directly in the action, leading to the reduced action $S_{\mathrm{dust}}$ in canonical form
\begin{equation} \label{actiond}
    S_{\mathrm{dust}} = \int dt \int dx \left[ 4 \pi \,\dot{T} \, p_T - N  \mathcal{H}^{(\mathrm{d})} -N^x  \mathcal{H}_x^{(\mathrm{d})} \right] \, ,
\end{equation}
where
\begin{subequations} \label{constraintsd}
    \begin{align}
        \mathcal{H}^{(\mathrm{d})}&=4\pi \,p_T\sqrt{1 +\frac{\abs{E^x}}{\left(E^\varphi\right)^2} \left(\partial_x T\right)^2}, \label{H_d} \\
        \mathcal{H}_x^{(\mathrm{d})} &= -4\pi\, p_T \,\partial_x T,  \label{Hx_d} 
    \end{align}
\end{subequations}
and the second-class constraints we solved for read
\begin{equation} \label{rho}
    P_\rho = 0 \quad\quad\quad\quad \mathrm{and} \quad\quad\quad\quad \rho = \frac{P_T}{\sqrt{q}} \bigg[\frac{\abs{E^x}}{\left(E^\varphi\right)^2} \left(\partial_x T\right)^2 +1\bigg]^{-1}.
\end{equation}
The total action $S$ of the system ``gravity + dust'' we are interested in is then given by
\begin{equation} \label{action}
    S=\int dt \int dx \left[ \frac{\dot{K}_x E^x + 2 \dot{K}_{\varphi} E^{\varphi}}{2 \gamma G} + 4 \pi \dot{T} \, p_T -N  \mathcal{H} - N^x  \mathcal{H}_x \right]  ,
\end{equation}
where 
\begin{equation}
    \mathcal{H} = \mathcal{H}^{(\mathrm{g})} + \mathcal{H}^{(\mathrm{d})},
\end{equation}
\begin{equation}
    \mathcal{H}_x = \mathcal{H}_x^{(\mathrm{g})} + \mathcal{H}_x^{(\mathrm{d})},
\end{equation}
whose exact expressions are given in \cref{constraintsg,constraintsd}, are respectively the Hamiltonian and the (radial) diffeomorphism constraints of the theory. The (generalized) Poisson brackets of our reduced phase space variables can be read directly from the action in \cref{action}, giving
\begin{subequations} \label{Poisson}
    \begin{align}
        &\{ K_x (x_1), E^x(x_2)\}= 2 \gamma G \,\delta(x_1-x_2) , \label{grav_x} \\
        &\{ K_{\varphi} (x_1), E^{\varphi}(x_2)\}= \gamma G \,\delta(x_1-x_2) , \label{grav_phi} \\
        &\{ T (x_1), p_T(x_2)\}= \frac{1}{4 \pi} \delta(x_1-x_2)  \label{dust} \, .
    \end{align}
\end{subequations}

\subsection{Gauge-fixing and reduction to single shell}
\label{sec:2.1}

Formally, this theory can be straightforwardly quantized using Dirac's quantization program~\cite{dirac2001lectures}. However, even in this symmetry-reduced setting, the analytic complexity of the constraints poses a serious obstacle to the explicit computation of the quantum dynamics. To gain insights into the quantum behavior of the theory, we will then fix as much gauge freedom as possible classically, so as to obtain a solvable quantum dynamics for this gauge-fixed theory. On a practical level, we need to find two gauge conditions that form a second-class set with our first-class constraints $\mathcal{H}$ and $\mathcal{H}_x$.

In the presence of matter, the natural choice for fixing the gauge is to tie the arbitrary choice of foliation (time function) and spatial coordinates in the ADM formalism with the matter degrees of freedom. As a first gauge-fixing condition, we choose the ``dust gauge'' condition $\chi_1 \coloneq T-t =0$. Namely, we choose the foliation naturally defined by the dust proper time field $T$ as the foliation used in the ADM decomposition. The Poisson brackets of this gauge condition with the first-class constraints are
\begin{subequations}
    \begin{align}
        \bigg\{\chi_1(x), \mathcal{H}(y) \bigg\} & \approx  \delta(x-y), \\
        \bigg\{\chi_1(x), \mathcal{H}_x(y) \bigg\} & \approx 0   ,
    \end{align}
\end{subequations}
where `$\approx$' stands for weak equality, that is equality up to expressions that vanish when constraints and gauge conditions are satisfied. So $\chi_1=0$ forms indeed a second-class pair with the Hamiltonian constraint $\mathcal{H}=0$, and it can then be used to gauge-fix the latter. This choice of gauge considerably simplifies the expressions of the dust part of the constraints in \cref{constraintsd}, giving
 \begin{equation} \label{constraintsd2}
    \mathcal{H}^{(\mathrm{d})}=4\pi p_T\quad\quad\quad\quad \mathrm{and} \quad\quad\quad\quad \mathcal{H}^{(\mathrm{d})}_x = 0.
\end{equation}
Finally, defining the total Hamiltonian $\mathcal H_{\mathrm{tot}}$ as
\begin{equation}
    \mathcal H_{\mathrm{tot}} = \int dx \Big[ N \mathcal{H} + N^x \mathcal{H}_x \Big],
\end{equation}
the conservation in time of the gauge condition, $\dot{\chi}_1 = 0$, gives
\begin{equation}
    \dot{\chi_1}=\frac{\partial \chi_1}{\partial t} + \{\chi_1 ,  \mathcal H_{\mathrm{tot}} \} \approx -1 + N  = 0,
\end{equation}
that is $N=1$.

The second gauge choice we want to impose is fixing the spatial coordinates on the leaves of the foliation to be comoving with the dust flow. In other words, we want the curves $x^a=\mathrm{const.}$ to be the trajectories of the dust fluid elements, and so $U^\mu \propto \partial^\mu_t$. Given the choice of foliation imposed by the dust gauge $\chi_1=0$, we see from \cref{ADM_lapse_shft} that this condition will be satisfied as long as $N^x=0$. Within the standard practice of considering $N$ and $N^x$ as Lagrange multipliers in the Hamiltonian theory, and not as phase variables, the condition $N^x=0$ is not a proper gauge condition, but it just fixes the value of the Lagrange multiplier $N^x$. However, if we instead consider $N^x$ and its conjugate momentum $\Pi_x$, satisfying the primary first-class constraint $\Pi_x=0$, as phase space variables, then $N^x=0$ is a proper gauge condition that forms a second-class set with $\Pi_x=0$. Both attitudes lead of course to the same reduced theory. The same applies also for the condition $N=1$ that follows from the conservation in time of the gauge condition $\chi_1=0$.

Solving the second-class constraints $\mathcal H = 0$ and $\chi_1=0$ for the pair $(T,p_T)$ and imposing the gauge-related conditions $N=1$ and $N^x=0$ inside the action, we obtain the reduced action
\begin{equation} \label{action_reduced1}
    S=\int dt \int dx \left[ \frac{\dot{K}_x E^x + 2 \dot{K}_{\varphi} E^{\varphi}}{2 \gamma G} -  \mathcal{H}^{(\mathrm{g})} \right] ,
\end{equation}
where $\mathcal{H}^{(\mathrm{g})}$ given in \cref{H_g} now serves as a true Hamiltonian for the system. Notice that while the diffeomorphism constraint $\mathcal{H}_x\equiv\mathcal{H}^{(\mathrm{g})}_x=0$ drops out of the action, it still has not been gauge-fixed. This simply means that while for each choice of initial conditions for our phase space variables $\big(K_x, K_\varphi, E_x,E_\varphi\big)$ there will be a unique dynamics, not all of these choices are gauge-independent, and so not all of them represent physically distinct scenarios. This is indeed consistent with the standard metric formulation of the LTB model in comoving and synchronous coordinates (i.e., in the comoving and dust gauges), where the line element takes the form
\begin{equation} \label{metricLTB}
    ds^2 = - dt^2 + \frac{(\partial_x r)^2}{1+\tilde{\varepsilon}} \, dx^2 + r^2\, d\Omega^2,
\end{equation}
with $r(t,x)$ denoting the areal radius of the shell $x$ at time $t$ and $\tilde{\varepsilon}(x)>-1$ being a parameter related to the total (Newtonian) mechanical energy of the shell $x$. The Einstein field equations read
\begin{equation} \label{EOMLTB}
    \dot{r}^2 = \frac{2Gm}{r} + \tilde{\varepsilon} \quad\quad\quad \quad \text{and} \quad\quad\quad \quad
    4\pi G \rho = \frac{ \partial_x m}{r^2\,\partial_x r},
\end{equation}
where $m(x)$ is the Misner-Sharp mass of spacetime at $x$. As we can see, the requirement of being in a comoving and synchronous coordinate system does not fix the radial coordinate. The metric in \cref{metricLTB} and the EOMs in \cref{EOMLTB} are in fact still invariant under an arbitrary reparametrization of $x$, leading to the gauge freedom generated by $\mathcal{H}_x$ in the Hamiltonian theory.

\Cref{action_reduced1} gives the canonical action for the LTB model in the comoving and dust gauges in terms of LQG elementary variables. Despite all the symmetry and gauge reduction, however, the expression for the true Hamiltonian $\mathcal{H}^{(\mathrm{g})}$, given in \cref{H_g}, is still too complex to provide a manageable dynamics in the quantum theory. Our last resort is then to gauge-fix the radial diffeomorphism constraint as well. Having tied as much gauge freedom as possible with the matter degrees of freedom, which are in fact no longer explicitly present in the reduced theory in \cref{action_reduced1}, we now turn to a purely metric condition. Following~\cite{Bojowald_2008,GieselEmbedding}, we can take advantage of the fact that we already know how the metric of the LTB model is supposed to look like, i.e. as in \cref{metricLTB}, and devise a gauge condition that is consistent with it. Consider in fact the so-called LTB condition\footnote{The condition in \cref{LTBcondition} is not actually the only possibility for an ``LTB condition''. See \cref{A} for more details.} 
\begin{equation} \label{LTBcondition}  
\chi_2 \coloneq E^{\varphi}-\frac{\partial_x E^x}{2 \sqrt{1+ \varepsilon}}=0,
\end{equation}
where $\varepsilon(x)$ is a given function of the radial coordinate $x$. The Poisson bracket of this condition with the radial diffeomorphism constraint reads
\begin{equation} \label{LTBPoisson}
        \big\{\chi_2 (x),  \mathcal{H}_x(y) \big\}  \approx -\left[E^\varphi(y)-\frac{\partial_y E^x(y)}{2 \sqrt{1+\varepsilon(x)}}\right] \partial_y \delta(x-y)  ,
\end{equation}
while its time evolution gives
\begin{equation} 
    \dot{\chi_2}= \{\chi_2 , \mathcal{H}^{(\mathrm{g})} \} \approx  0 \, .
\end{equation}
For $\varepsilon \neq 0$, the LTB condition in \cref{LTBcondition} is a proper gauge-fixing condition for the diffeomorphism constraint that is also automatically conserved in time, and so it does not lead to any further constraints. However, we are interested in the simpler marginally bound case $\varepsilon=0$. In this case, as already noticed in~\cite{GieselGeneralized,GieselEmbedding}, the right-hand side of \cref{LTBPoisson} vanishes, and the LTB condition cannot be used to gauge-fix the radial diffeomorphism constraint.

The LTB condition effectively constrains the line element in \cref{metricE} to take the LTB form given in \cref{metricLTB}, both in the marginally bound and non-marginally bound cases. However, it is only in the latter case that the condition also serves as a gauge-fixing of the diffeomorphism constraint.  The reason is that, for $\varepsilon\neq0$, the LTB condition not only restricts the form of the line element, but it also constrains $x$ to be the specific radial coordinate in which $\tilde{\varepsilon}(x)$ assumes the given expression $\varepsilon(x)$ entering \cref{LTBcondition}. This second property is what makes the LTB condition, at least locally, a valid gauge condition for the diffeomorphism constraint in the non-marginally bound case. In contrast, this property is completely lost for $\varepsilon=0$. As a consequence, in the marginally bound case, the LTB condition does not act as a gauge condition but as a further first-class constraint.\footnote{\label{footnote}It is furthermore well known, but often not emphasized enough, that proper gauge conditions for first-class constraints generating spacetime diffeomorphisms must depend explicitly on the spacetime coordinates~\cite{Barros_e_Sa_2025}. Thus, even if the Poisson bracket in \cref{LTBPoisson} did not vanish for $\varepsilon=0$, the marginally bound LTB condition would still fail to be a proper gauge condition for the radial diffeomorphism constraint, as it lacks an explicit dependence on $x$.} If we were to consider the ``new'' system defined by the action in \cref{action_reduced1} with the addition of the marginally bound LTB condition, the resulting theory would possess four phase space variables $\big(K_x, K_\varphi, E_x, E_\varphi\big)$ and two first-class constraints, and therefore no local degrees of freedom.

This outcome is not surprising, as it is perfectly consistent with the analogous analysis carried out in the Lagrangian formulation of the theory. In the marginally bound case $\varepsilon=0$, the only role of the LTB condition is to restrict the line element in \cref{metricE} to the LTB form in \cref{metricLTB}. In other words, the condition effectively reduces the number of degrees of freedom of the theory by exploiting a known symmetry of the system --- in this case, the specific form of the line element once Einstein’s field equations have already been solved. While such a reduction is always possible at the level of the equations of motion, it may fail at the level of the action. This limitation is evident in the marginally bound LTB model of \cref{metricLTB,EOMLTB}, even before carrying out any explicit calculation. When $\varepsilon=0$, the metric contains only one degree of freedom, the areal radius $r(t,x)$. Substituting the reduced metric in \cref{metricLTB} directly into the Einstein field equations yields the two equations of motion in \cref{EOMLTB}. However, inserting the same reduced metric into the action in metric variables, there is no way to recover both equations of motion from an action that depends only on the single variable $r$.

The explicit computation in the Lagrangian formulation can be found in~\cite{Kiefer_2019}. There, the marginally bound LTB metric ansatz is inserted into the action in metric variables, and it is shown that the action reduces to a boundary term in the radial coordinate $x$.\footnote{Although this result is obtained in~\cite{Kiefer_2019} using both the LTB metric ansatz and the equations of motion, it can in fact be obtained from the LTB metric ansatz alone.} The resulting Euler-Lagrange equations reproduce only the first equation in \cref{EOMLTB}, evaluated at the chosen boundary value of $x$. The second original equation of motion, which determines the matter energy density $\rho$ in terms of the metric degrees of freedom, and therefore encodes the relative dynamics of neighboring shells, is lost in this reduction process. As a consequence, the reduced action correctly describes the dynamics of an individual shell (indeed, of any given shell) of the collapsing matter, but fails to reproduce the collective behavior of the collapsing matter as a whole. In agreement with our Hamiltonian analysis, the theory possesses no local degrees of freedom.

This reduced theory was used in~\cite{Kiefer_2019} to study the quantum dynamics of a single shell of the marginally bound LTB model in metric variables, and to construct an effective quantum-corrected metric for the entire LTB spacetime. A significant drawback of this approach is that, in order to make statements about the entire collapsing dust cloud, it is necessary to assume that the shells can be treated independently from each other. While this is always true classically (in the absence of shell-crossing singularities), it does not necessarily hold in the quantum theory. Although this is a serious limitation, the simplicity of this single-shell model makes it possible to perform a detailed quantum analysis and to compute its dynamics explicitly, something that would be otherwise unfeasible. For this reason, we begin our analysis by investigating the loop quantization of this reduced model and comparing it with the metric quantization presented in~\cite{Kiefer_2019}.

The marginally bound LTB condition can be imposed in the Hamiltonian theory by adding the constraint $\chi_2$ to the action in \cref{action_reduced1} using a new Lagrange multiplier $\lambda$:
\begin{equation} \label{action_reduced2}
    S=\int dt \int dx \left[ \frac{\dot{K}_x E^x + 2 \dot{K}_{\varphi} E^{\varphi}}{2 \gamma G} -  \mathcal{H}^{(\mathrm{g})} - \lambda \chi_2 \right] ,
\end{equation}
where the diffeomorphism constraint $\mathcal{H}_x\equiv\mathcal{H}^{(\mathrm{g})}_x=0$ does not enter the action but still needs to be fixed. In order to completely fix the gauge, we then require two gauge-fixing conditions $C_1=0$ and $C_2=0$ that form a second-class set with $\chi_2=0$ and $\mathcal{H}_x=0$. A suitable choice is given by
\begin{align}
    C_1 &=E^x K_x - E^\varphi K_\varphi = 0, \\
    C_2 &=K_\varphi^2 \sqrt{\abs{E^x}}- F(x) = 0,
\end{align}
where $F(x)$ is a given function of $x$ (see \cref{footnote}), whose explicit form is not relevant for what follows. It is straightforward to show that these conditions are second class with respect to $\chi_2=0$ and $\mathcal{H}_x=0$, and their conservation in time gives
\begin{align}
    \dot{C_1}&=\{C_1, \int dx  \, \left( \mathcal{H}^{(\mathrm{g})} + \lambda \chi_2 \right) \} \approx E^x \partial_x\lambda-\frac{1}{2}\partial_x E^x \lambda\ = 0,\\
    \dot{C_2}&=\{C_2, \int dx  \, \left( \mathcal{H}^{(\mathrm{g})} + \lambda \chi_2 \right) \} \approx 2 \gamma G \lambda K_\varphi \sqrt{\abs{E^x}} = 0\, .
\end{align}
The second equation gives us $\lambda=0$, which is enough to also satisfy the first equation. Thus, there are no further constraints to consider. We can then resolve all the second-class constraints in the action to obtain
\begin{equation} \label{action_reduced3}
    S=\frac{1}{2 \gamma G} \int dt \int dx \;  \partial_x \left( \dot{K}_{\varphi} E^x +  \sgn(E^\varphi) \sgn(E^x) \frac{K_{\varphi}^2 \sqrt{\abs{E^x}}}{\gamma} \right) ,
\end{equation}
where $\sgn(E^\varphi)=\sgn(g)$ and $\sgn(E^x)=\sgn(f)$. A priori, the relative sign between $E^\varphi$ and $E^x$ is not fixed by the LTB condition, which gives $\sgn(E^\varphi)=\sgn(\partial_x E^x)$. Namely, it implies that for positive (negative) $E^\varphi$, $E^x$ must be an increasing (decreasing) function of $x$, but it gives no information on the sign of $E^x$ itself. However, regularity at the center $x_{\mathrm c}$ of the dust cloud requires the square of the areal radius of the center, $\abs{E^x (x_{\mathrm c})}$, to vanish. This further condition can be satisfied only if $\sgn(E^\varphi)=\sgn( E^x)$.

We have thus found that the reduced action of the system is a boundary term in $x$, in agreement with the counting of degrees of freedom and the Lagrangian theory. If we now take as boundary of the integration in the radial coordinate the center of the dust cloud $x_{\mathrm c}$ and an arbitrary shell $x>x_{\mathrm c}$ in the cloud, we obtain
\begin{equation} \label{action_reduced4}
    S= \int dt    \Bigg[ \frac{\dot{K}_{\varphi}(x) E^x(x) }{2 \gamma G}- H(x) \Bigg] ,
\end{equation}
where
\begin{equation} \label{Ham}
    H(x)\coloneq -\frac{K^2_{\varphi}(x) \sqrt{\abs{E^x (x)}}}{2 \gamma^2 G} 
\end{equation}
and we used the fact that the square of the areal radius $\abs{E^x (x_{\mathrm c})}$ of the center of the dust cloud must vanish. This action, together with its associated non-field-theoretic Hamiltonian $H(x)$, leads to non-distributional brackets 
\begin{equation} \label{Pbrackets}
    \{K_{\varphi}, E^x\}(x)=2 \gamma G
\end{equation}
and provides the canonical description of the dynamics of any given shell $x$ in a dust cloud. Notice that, after the gauge-fixing, the energy density $\rho$ in \cref{rho} reads
\begin{equation}
    \rho(x) = - \frac{\partial_x H(x)}{4\pi \sqrt{q}} \sin \theta.
\end{equation}
The active gravitational mass $m(x)$ of spacetime (in the marginally bound case) at the radial coordinate $x$ is then given by
\begin{equation} \label{H=-E}
    m(x) = \int d^3 \tilde{x} \,\sqrt{q} \, \rho(\tilde{x}) = - H(x).
\end{equation}
Namely, the Hamiltonian $H(x)$ controlling the dynamics of the dust shell $x$ is equal to the negative total mass-energy content of the ball of dust enclosed in the shell $x$. The EOMs for the shell $x$ are given by
\begin{subequations} \label{EOM}
    \begin{align}
        &\dot{K}_{\varphi} (x) =\{K_{\varphi}, H\} (x) =- \frac{\sgn(E^x)}{2 \gamma} \frac{K_{\varphi}^2(x)}{\sqrt{\abs{E^x(x)}}},  \label{K_dot} \\
        &\dot{E}^x (x) =\{E^x, H\} (x) = \frac{2}{\gamma} K_{\varphi}(x) \sqrt{\abs{E^x(x)}}  \label{E_dot} \, .
    \end{align}
\end{subequations}
It is straightforward to check that these equations are consistent with the first equation in \cref{EOMLTB}, thus giving the correct dynamics for each shell $x$. Moreover, given that the LTB coordinates used here also cover the vacuum region in the exterior of the collapsing matter, the same Hamiltonian could be used to describe vacuum shells as well. However, in that case, one finds $H(x>\tilde{x})=-M$, where $\tilde{x}$ is the radial coordinate of the boundary of the collapsing matter and $M$ is a constant representing its total mass-energy.

This completes the analysis of the classical theory. Before turning to the quantum theory, it is worth noting that the Hamiltonian in \cref{Ham} is negative and has no lower bound. The possibility of a negative and unbounded Hamiltonian in a general relativistic setting is a consequence of the non-positive definiteness of the ``kinetic term'' in the Einstein-Hilbert action~\cite{DeWitt}, and is linked to the attractive nature of gravity~\cite{Giulini_1994}. In quantum mechanics, one typically works with Hamiltonians that are bounded from below, as these possess a well-defined ground state. This is because the Hamiltonian usually coincides with the energy of the system, and the absence of a ground state for the energy would lead to an unstable theory. For this reason, in~\cite{Giesel_2010} --- where dust is used as a matter reference frame for other matter fields as well as for the gravitational field --- it is argued that the dust energy density $\rho$ should be taken to be negative in order to obtain a positive-definite Hamiltonian. However, as clearly shown by \cref{H=-E}, the Hamiltonian of our reduced system is not an energy, but rather its negative. Therefore, even though the Hamiltonian is negative, the energy remains positive and the theory is well defined.

\section{Quantum theory} \label{sec:3}

At this point, we could just promote the classical variables to operators, their Poisson brackets to commutators, and use the Hamiltonian in \cref{Ham} to compute the dynamics of quantum states associated with each shell. This would lead to the standard Wheeler-DeWitt quantization of the symmetry-reduced system, as carried out in~\cite{Kiefer_2019}. Instead, our aim is to quantize the system in a way that follows as closely as possible the loop quantum gravity framework, as in loop quantum cosmology.

In loop quantum gravity, the elementary variables are the holonomies $h_\alpha (A)$ of the Ashtekar-Barbero connection $A_a^i$ along curves $\alpha$ and the fluxes $E(\mathcal{S})$ of the densitized triad $E^a_i$ across surfaces $\mathcal S$. After the gauge reduction, the only remaining degree of freedom of the densitized triad in our model is $E^x$. Furthermore, due to spherical symmetry, its flux across any surface lying on a fixed shell $x$ is simply proportional to $E^x$, with the proportionality factor depending only on the specific surface $\mathcal{S}$. As in loop quantum cosmology~\cite{Ashtekar:2003hd}, we may therefore take $E^x$ itself as the flux elementary variable of the symmetry-reduced model. Spherical symmetry also implies that not all curves $\alpha$ on a fixed shell $x$ are necessary to (classically) reconstruct the connection from the holonomies $h_\alpha (A)$. We may thus restrict attention to curves tangent to $\partial_\theta$ at constant $\varphi$.\footnote{One could consider directly ``point holonomies'', namely complex exponentials of the non-integrated connection. We instead consider holonomies along curves $\varphi=\mathrm{const.}$ in order to maintain closer contact with the heuristics underlying the $\bar{\mu}$-scheme.}

As we will show in \cref{sec:3.2}, these elementary variables lead to a well-defined quantum kinematics for our model. However, as in full loop quantum gravity, there is no well-defined operator corresponding to the connection $A$ itself in this quantum representation. This poses a difficulty because the Hamiltonian $H(x)$ in \cref{Ham} depends explicitly on $K_\varphi$, which is the only remaining dynamical degree of freedom of the connection.\footnote{Notice however from \cref{connection} that $K_\varphi$ is not directly a component of $A$.} It is therefore necessary to rewrite the Hamiltonian in terms of the new elementary variables, i.e., holonomies, and then quantize this new expression. In loop quantum cosmology, the standard approach is to mirror the regularization of the Hamiltonian constraint in full LQG by expressing the field strength $F$ of the Ashtekar-Barbero connection in terms of holonomies around closed loops (also known as plaquettes), in close analogy with lattice gauge theory. In this way, the symmetry-reduced Hamiltonian constraint can be written classically as the limit for the area of the loops that goes to zero of an expression involving only holonomies and fluxes. For any fixed nonzero area of the loops, this expression admits a well-defined operator representation in the quantum theory. The limit of vanishing loop area, however, does not exist. Rather than being a shortcoming, this feature reflects a fundamental physical aspect of the theory: in a quantum theory where geometry itself is quantized, as in loop quantum gravity, the local notion of connection (or field strength) loses any meaning, and no operator corresponding to it can be defined. The best possible approximation is then obtained by choosing the physical area of the plaquettes to be the smallest non-zero eigenvalue $\Delta= 4 \sqrt{3} \pi \gamma \hbar G $ of the area operator in the full theory. This prescription~\cite{Ashtekar_2006_2}, commonly known as the $\bar{\mu}$-scheme, provides a powerful heuristic for constructing a well-defined Hamiltonian (constraint) operator in symmetry-reduced models.

In models with non-vanishing spatial curvature, of which spherically symmetric systems are a generic example, this construction is not always available. Indeed, in some of these cases, there exists no well-defined operator on the kinematical Hilbert space defined by the holonomy-flux algebra that represents the holonomy of the Ashtekar-Barbero connection around a closed loop~\cite{Ashtekar:2009um}. We are then led to abandon the regularization based on the field strength $F$, and instead look for a direct regularization of the connection $A$ itself. Since the fundamental variables of loop quantum gravity are holonomies of the Ashtekar-Barbero connection, a natural strategy from the perspective of the full theory is to express $A$ in terms of its holonomies and then quantize this expression. However, this procedure also fails to yield a well-defined operator in certain cases~\cite{Vandersloot:2006ws}. The solution, known as ``$K$-quantization''~\cite{Vandersloot:2006ws,Singh:2013ava}, is to consider holonomies of the extrinsic curvature $K$, rather than of $A$, as the fundamental variables. It can be shown~\cite{Singh:2013ava} that this prescription always leads to well-defined quantum operators in such settings, and that, whenever the field strength prescription is also available, the two approaches give compatible results.\footnote{By contrast, there are cases in which both the field strength prescription and the A-prescription are available, yet they lead to non-compatible results.~\cite{Singh:2013ava}}

In our model, the only dynamical degree of freedom of the connection is the extrinsic curvature component $K_{\varphi}$, which also appears explicitly in the Hamiltonian $H(x)$. It is therefore natural to adopt a direct quantization of the extrinsic curvature, as prescribed by the $K$-quantization scheme. We are thus led to consider holonomies $h_{\mu}(x)$ of the extrinsic curvature $\gamma K$, with the Barbero-Immirzi parameter included for notational simplicity, along curves tangent to $\partial_\theta$, that is, along meridian arcs:
\begin{equation} \label{Kholonomy}
    \begin{split}
    h_{\mu} (x)& = \exp \left(\int_0^\mu \gamma K_\theta^i (x) \tau_i \, d \theta \right)  \\
    &=\exp \left( \int_0^\mu K_\varphi (x) \tau_2 \, d \theta\right) \\
    &= \cos\left(\frac{\mu K_\varphi (x)}{2}\right) \mathbbm{1} + 2  \sin\left(\frac{\mu K_\varphi (x)}{2}\right) \tau_2 \, ,
\end{split}
\end{equation}
where $\mu$ is the length of the chosen curve, i.e., the angle spanned by the arc. The elementary variables of the theory may then be taken to be the point holonomies
\begin{equation} \label{Nholonomy}
    \mathcal{N}_\mu (x) = \exp\left( \frac{i \mu K_\varphi (x)}{2} \right) ,
\end{equation}
which, together with the fluxes $E^x(x)$, generate the algebra of elementary variables, with Poisson brackets
\begin{equation}
    \big\{\mathcal{N}_\mu,E^x\big\}(x) = i\gamma G\,\mu  \,\mathcal{N}_\mu (x).
\end{equation}
Although this algebra is classically equivalent to the one generated by $K_\varphi$ and $E^x$, it leads to a unitarily inequivalent representation in the quantum theory. The extrinsic curvature component $K_{\varphi}$ can be expressed in terms of these variables as
\begin{equation} \label{Kinholonomy}
    K_{\varphi} (x) = \lim_{\mu\rightarrow 0} \frac{ \mathcal{N}_{2\mu }(x)- \mathcal{N}_{-2\mu} (x)}{2i\mu} \,.
\end{equation}
This relation provides the desired expression of $K_{\varphi}$ in terms of its holonomies.

In close analogy with the $F$-prescription in loop quantum cosmology, for any fixed nonzero value of $\mu$, the expression on the right-hand side of \cref{Kinholonomy} can be promoted to a well-defined operator in the quantum theory constructed in \cref{sec:3.2}. Once again, however, the limit for $\mu\rightarrow0$ of these operators does not exist. As already mentioned, this is a physical consequence of the underlying discrete quantum geometry: no operator corresponding exactly to $K_{\varphi}$ can be defined in the quantum theory. The best one can do is to approximate $K_{\varphi}$ by evaluating the right-hand side of \cref{Kinholonomy} at the smallest fixed (but not constant) value of $\mu$ allowed by the discrete quantum geometry, in direct analogy with the $\bar{\mu}$-scheme. One way to determine this value is to compare the $F$- and $K$-prescriptions in models where both are available, and to fix the value of $\mu$ in the $K$-prescription by requiring that the resulting field strength operator agree with the field strength operator of the $\bar{\mu}$-scheme in the $F$-prescription~\cite{Ashtekar:2009um}. This procedure leads to
\begin{equation} \label{mubar}
    \mu \longrightarrow \bar{\mu} \coloneq \sqrt{\frac{\Delta}{\abs{E^x}}}\,,
\end{equation}
where $\bar\mu$ is, crucially, phase space dependent. Equivalently, one may mimic the plaquette discretization of the $F$-prescription using holonomies of the extrinsic curvature in the $K$-prescription, as done in~\cite{Chiou:2008eg}. Requiring the physical area of these plaquettes to equal $\Delta$ again yields \cref{mubar}. It is also worth noting that $\bar{\mu}$ coincides with the value of $\mu$ for which the physical length $\delta l$ of the arc associated with the holonomy in \cref{Kholonomy} is precisely given by $\delta l = \bar{\mu} \sqrt{\abs{E^x}} = \sqrt{\Delta}$.

Let $ \mathcal{N}(x)\coloneq \mathcal{N}_{\bar{\mu}} (x) $. We have thus obtained that, in the quantum theory, the extrinsic curvature component $K_{\varphi}$ should be approximated as
\begin{equation} \label{K_to_sin}
   K_\varphi \longrightarrow  \lim_{\mu\rightarrow \bar{\mu}} \frac{ \mathcal{N}_{2\mu }(x)- \mathcal{N}_{-2\mu} (x)}{2i\mu}=\frac{\mathcal{N}{\!\,\!\,}^2 (x) - \mathcal{N}^{-2}(x)}{2i\bar\mu} \, .
\end{equation}
The action of this quantity in the quantum theory takes a particularly simple form when the phase space is reparametrized in terms of variables adapted to $ \mathcal{N}(x)$, namely
\begin{align}
    v (x) &\coloneqq \frac{2}{3} \frac{\abs{E^x (x)}^{\frac{3}{2}}}{\gamma \sqrt{\Delta} G \hbar} \sgn\left(E^x\right) ,\\
    b (x)& \coloneqq  \sqrt{\Delta} \frac{K_\varphi (x)}{\sqrt{\abs{E^x}}} = \bar{\mu} (x) K_\varphi (x)\, ,
\end{align}
which satisfy
\begin{equation} \label{Pbracket2}
    \{b, v\} (x)=\frac{2}{\hbar}.
\end{equation}
Here, $b(x)$ is precisely the phase space variable appearing in the point holonomies $\mathcal N(x)$, while $v(x)$ is its conjugate variable. Incidentally, $v(x)$ is also related to the physical volume $V(x)= 4 \pi \abs{E^x(x)}^{\frac{3}{2}}/3$ enclosed by a shell $x$ via 
\begin{equation} \label{Vv}
    V(x)=2 \pi  G\gamma \sqrt{\Delta}  \hbar\, \abs{v(x)}. 
\end{equation}
Expressed in these variables, the physical Hamiltonian $H(x)$ in \cref{Ham} becomes
\begin{equation} \label{H_bv}
    H(x)=-\frac{3}{4} \frac{\hbar}{\gamma \sqrt{\Delta}}\, b^2(x) \abs{v(x)} .
\end{equation}
It follows immediately that $b(x)$ is related to the global density $\rho{}_{{}_{\mathrm G}} (x)$ of total mass-energy enclosed by the shell $x$ through
\begin{equation} \label{rhoG}
    \rho{}_{{}_{\mathrm G}} (x) \coloneq \frac{m(x)}{V(x)} = -\frac{H(x)}{V(x)} = \rho_{\mathrm c} \,b^2 (x),
\end{equation}
where $\rho_{\mathrm c} = 3/(8\pi G\gamma^2\Delta)$, known as the critical density, will play a central role in the quantum theory. Finally, the Poisson brackets between $\mathcal N (x) \coloneq \exp \big(i b(x)/2\big)$ and $v(x)$ read
\begin{equation} \label{Pbracket3}
    \big\{\mathcal{N},v\big\}(x) = \frac{i}{\hbar}\mathcal{N} (x).
\end{equation}

For each fixed shell $x$, the expression in \cref{H_bv} is mathematically identical (up to constant factors) to the Hamiltonian describing a spatially-flat Friedmann universe filled with dust~\cite{Kiefer_2019,Husain:2011tm}. As a consequence, the quantum theory of a single shell in our model leads to the same physical results as the corresponding cosmological model. Interestingly, the dust-filled cosmological scenario has received little attention in the LQC literature, which has instead focused primarily on the case of a massless scalar field~\cite{Ashtekar:2011ni}. To the best of the authors’ knowledge, \cite{Husain:2011tm} is the only reference addressing a spatially-flat Friedmann universe filled with dust in the $\bar\mu$-scheme,\footnote{Its $\mu_0$-scheme is briefly discussed in~\cite{Willis:2004br}} and it does not include a numerical analysis of the model's dynamics. For these reasons, and in order to keep the present work self-contained, we review the construction of the quantum theory associated with the algebra in \cref{Pbracket3} and perform a numerical investigation of the dynamics generated by the Hamiltonian in \cref{H_bv}.

In the following subsections, we analyze the Wheeler-DeWitt quantization of the single-shell model in the $b$-$v$ algebra, then investigate the loop quantization of the holonomy-flux algebra constructed above, and lastly construct a quantum multi-shell model from the single-shell one.

\subsection{Wheeler-DeWitt quantum theory} \label{sec:3.1}

We now explore the canonical quantization of the single-shell system in the $b$-$v$ algebra. We use these variables, rather than $K_\varphi$ and $E^x$ or the original metric variables, solely to facilitate the comparison with the loop quantization discussed in the next subsection, which is based on the $b$-$v$ holonomy flux algebra in \cref{Pbracket3}. Since working in the $b$-$v$, $K_\varphi$-$E^x$, or metric-variable representations leads to the same physical picture --- unlike the unitarily inequivalent loop representation --- we refer to this quantization as the Wheeler-DeWitt (or geometrodynamical) quantization, following the standard terminology for canonical quantization in metric variables.

Having classically gauge-fixed all the constraints of the theory, we are left with a symmetry-reduced model whose dynamics is given by a true physical Hamiltonian. As a result, the canonical quantization proceeds exactly as in standard non-relativistic quantum mechanics. For each shell $x$, the quantum Hilbert space is $\underline{\mathscr{H}}^{(x)} = L^2(\mathbb{R}, dv)$ equipped with the scalar product
\begin{equation}
    \underline{ \braket{\varphi_1}{\varphi_2}}=\int_{-\infty}^{+\infty} dv \, \varphi_1^*(v)  \, \varphi_2(v) \, .
\end{equation}
The classically conjugated variables $b(x)$ and $v(x)$ are promoted to operators $\underline{\hat{b}}_x$ and $\underline{\hat{v}}_x$ acting on $\underline{\mathscr{H}}^{(x)}$, satisfying the commutation relation $\big[\, \underline{\hat{b}}_x, \underline{\hat{v}}_x \big]= 2i$ (see \cref{Pbracket2}). Following standard conventions in the loop quantum cosmology literature, underbars are used to distinguish WDW quantities from their LQG counterparts. From now on, for notational simplicity, we suppress the explicit dependence of the operators on $x$. In this $v$-representation, the operators $\underline{\hat{v}}$ and $\underline{\hat{b}}$ take the familiar multiplicative and differential forms,
\begin{align}
    &\underline{\hat{v}} \,  \varphi(v) = v \,\varphi (v), \\
    &\underline{\hat{b}} \,  \varphi(v) = 2 i \frac{d}{dv}  \varphi(v) \label{b_WDW} \, .
\end{align}

Upon choosing an appropriate factor ordering, the Hamiltonian $H$ in \cref{H_bv} is promoted to an operator $\underline{\hat{H}}$, and the quantum dynamics is dictated by a Schrödinger equation describing the evolution of states in time $T$.\footnote{In the quantum theory, evolution proceeds with respect to the time $t$ appearing in the action in \cref{action_reduced4}. However, to emphasize that this time corresponds to the dust proper time fixed by the dust gauge $t=T$, we use the dust field $T$ as the time parameter in the quantum theory.} In principle, one could consider a two-parameter family of factor orderings for $\underline{\hat{H}}$, as discussed in~\cite{Kiefer_2019}. In what follows, however, we restrict attention to the same factor ordering employed in the loop quantization, in order to facilitate a direct comparison between the two approaches. This choice corresponds to the so-called MMO factor ordering~\cite{MMO,Prescriptions} in the LQC literature. With this prescription, the Hamiltonian operator $\underline{\hat{H}}$ takes the form
\begin{equation} \label{H_WDW}
    \underline{\hat{H}}  \coloneq - \frac{3 \hbar}{4 \gamma \sqrt{\Delta}} \abs{\underline{\hat{v}}}^{\frac{1}{4}} \underline{\hat{D}} \,\abs{\underline{\hat{v}}}^{\frac{1}{2}} \underline{\hat{D}} \,\abs{\underline{\hat{v}}}^{\frac{1}{4}}  ,
\end{equation}
where $\underline{\hat{D}}\coloneq\frac{1}{2}\left( \sgn(\underline{\hat{v}})\underline{\hat{b}}+\underline{\hat{b}} \sgn(\underline{\hat{v}}) \right)$.

The action of the Hamiltonian $\underline{\hat{H}}$ in $\underline{\mathscr{H}}^{(x)}$ is explicitly given by
\begin{equation} \label{H_WDW2}
    \underline{\hat{H}} \varphi (v) =\frac{3 \hbar}{\gamma \sqrt{\Delta}} \left( -\frac{1}{16}\frac{1}{\abs{v}} \varphi ( v) + \sgn(v) \varphi^\prime(v) + \abs{v} \varphi^{\prime \prime}( v) \right),
\end{equation}
where $\varphi^\prime \equiv \partial_v \varphi$. In deriving this expression, we have already imposed the regularity condition that the term proportional to $\delta(v) \varphi(v)$ be non-singular, which requires $\varphi(0)=0$. For the action of $\underline{\hat{H}}$ to be well defined, the right-hand side of \cref{H_WDW2} must exist and belong to $L^2(\mathbb{R},v)$. This condition is satisfied in $v=0$ if the quantity $\abs{v}^{\frac{1}{2}}\partial_v \left( \abs{v}^{\frac{1}{4}} \varphi(v)\right)$ is differentiable, an assumption we adopt from now on. It is also worth noting that the parity operator $\hat{\Pi} \varphi(v) = \varphi(-v)$, which implements a reversal of the triad orientation, commutes with $\underline{\hat{H}}$. Parity of the triad orientation is therefore a symmetry of both the classical and the quantum theory. As a consequence, the Hilbert space $\underline{\mathscr{H}}^{(x)}$ decomposes into the even and odd eigenspaces of $\hat{\Pi}$, and the Hamiltonian preserves these two sectors. Moreover, since we are not interested in observables that are sensitive to the orientation of the triad --- just as in the Wheeler–DeWitt theory formulated in metric variables --- the parity operator commutes with all the observables of interest: although $\hat{\Pi}$ does not commute with $\underline{\hat{v}}$ and $\underline{\hat{b}}$, which are by construction sensitive to the triad orientation, it does commute e.g. with $\underline{\hat{V}} \propto\abs{\underline{\hat{v}}}$ and $\underline{\hat{\rho}}{}_{{}_{\mathrm G}} \propto \underline{\hat{b}}^2$ (see \cref{Vv,rhoG}). Its eigenspaces therefore define superselection sectors. Without loss of generality, we restrict attention to the symmetric sector. All in all, the domain of $\underline{\hat{H}}$ can be taken to be
\begin{equation} \label{DH}
    D(\underline{\hat{H}})=\Big\{\varphi \in \mathcal{S}(\mathbb{R}):  \varphi(0)=0  ,\;  \; \varphi(v)=\varphi(-v),  \; \; \abs{v}^{\frac{1}{2}}\partial_v \left( \abs{v}^{\frac{1}{4}} \varphi(v)\right) \in C^1(\mathbb{R}) \Big\}
\end{equation}
where $\mathcal{S}(\mathbb{R})$ denotes the Schwartz space.


The spectrum of $\underline{\hat{H}}$ is purely absolutely continuous\footnote{Even if a point spectrum were present, it would not contribute to the physical description of gravitational collapse, just as bound states do not contribute to scattering states in a quantum mechanical system with a potential well.} and given by $\mathrm{Sp}\big(\underline{\hat{H}}\big)=(-\infty,0]$. Its generalized eigenstates $\underline{\mathrm e}_{\pm k}$ with $k\in\mathbb{R}^+$ --- excluding the zero-eigenvalue solution, which, as in the case of the free particle Hamiltonian, does not correspond to a physical state --- satisfy 
\begin{equation} \label{ek_WDW}
    \underline{\hat{H}} \,\underline{\mathrm e}_{\pm k} = -\omega_k^2 \,\underline{\mathrm e}_{\pm k} \quad\quad \text{with}\quad\quad 
    \omega_k^2 =\frac{3\hbar}{4\gamma\sqrt{\Delta}} k^2 > 0,
\end{equation}
are two-fold degenerate, and take the form\footnote{Although the complex exponential eigenstates $\underline{\mathrm e}_{\pm k} (v)$ are not normalizable in the Dirac-delta sense, they can be combined to form sine and cosine eigenfunctions that admit delta-function normalization.}
\begin{equation}
    \underline{\mathrm e}_{\pm k} (v)= \frac{1}{\sqrt{4\pi}\abs{v}^{1/4}}\, e^{\pm i k \sqrt{\abs{v}}},
\end{equation}
Unlike the typical situation in non-relativistic quantum mechanics, the Hamiltonian in this model is negative definite and unbounded from below. However, as already discussed at the end of \cref{sec:2}, the Hamiltonian here corresponds to the negative of the energy. For this reason, the energy operator is positive definite, and the quantum theory is well defined.

A direct analysis, outlined in \cref{B}, shows that although $\underline{\hat{H}}$ is not self-adjoint, it admits a one-parameter family of self-adjoint extensions $\underline{\hat{H}}_a$, defined on the domains
\begin{equation}  \label{DHa}
        D(\underline{\hat{H}}_{a})=\Big\{\varphi(v) \in \mathcal{S}(\mathbb{R}) : \;\varphi(v)=\varphi(-v),\;\; \lim_{\;\,v\rightarrow 0^+}\left[v^\frac{1}{2} \partial_v \left(v^\frac{1}{4} \varphi(v)\right) - a \, v^\frac{1}{4}\varphi(v)\right]=0 \Big\} ,
\end{equation}
where $a\in\mathbb{R} \,\cup\,\{\infty\}$. The generalized eigenfunctions $ \underline{\mathrm e}_{a, \, k}$ of $\underline{\hat{H}}_a$ can be expressed as (see \cref{eakWDW})
\begin{equation}
    \underline{\mathrm e}_{a, \, k} (v) = \frac{1}{\sqrt{2 \pi}\abs{v}^{\frac{1}{4}}}  \cos\left(k \sqrt{\abs{v}} + \frac{\beta(a,k)}{2} \right),
\end{equation}
where $\beta(a,k)$ is implicitly defined by $a=-(k/2) \tan\left(\beta/2\right)$. For each choice of $a$, the Schrödinger equation
\begin{equation} \label{Schr_WDW}
    i \hbar \frac{\partial }{\partial T} \varphi (T, v) = \underline{\hat{H}}_a \, \varphi (T, v) \, 
\end{equation}
generates a unitary time evolution via the operator $\hat{U}_a (T)\coloneq \exp\big(-iT\underline{\hat{H}}_a/\hbar\big)$. Each self-adjoint extension $\underline{\hat{H}}_a$ therefore defines a physically distinct quantum dynamics. Given an arbitrary initial state
\begin{equation}
     \varphi(T=0, v) = \int_{0}^{+\infty} dk \, \,\tilde{\varphi} (k) \, \underline{\mathrm e}_{a, \, k} (v)
\end{equation}
with some spectral profile $\tilde{\varphi} (k) $, its time evolution under the dynamics generated by $\underline{\hat{H}}_a$ is given by
\begin{equation} \label{physStatesWDW}
    \varphi (T, v)= \int_{0}^{+\infty} dk \, \,\tilde{\varphi} (k) \, \underline{\mathrm e}_{a, \, k} (v)
     e^{i\omega_k^2 T/\hbar}.
\end{equation}

A natural question at this stage is whether a physically motivated criterion exists that would single out a preferred value of the extension parameter $a$. DeWitt~\cite{DeWitt}, motivated by the presence of the classical singularity at $v=0$ (that is, $V=0$, see \cref{Vv}), argued for selecting $a$ so as to ensure singularity avoidance in the quantum theory. He therefore proposed imposing the condition $\varphi(v=0)=0$.\footnote{Notice that while $\varphi(0)=0$ always holds for $\varphi \in D(\underline{\hat{H}})$ (see \cref{DH}), this is no longer true for all $\varphi \in D(\underline{\hat{H}}_a)$ (see \cref{DHa}).} Assuming the standard probability interpretation of $\abs{\varphi}^2$, which itself raises conceptual issues in quantum gravity, this condition implies a vanishing probability for the occurrence of a singular collapsed state in the quantum theory.

This criterion, however, is inadequate. From a technical standpoint, its applicability depends crucially on the choice of factor ordering for $\underline{\hat{H}}$. For some orderings, the condition $\varphi(0)=0$ uniquely fixes the value of $a$, whereas for others --- such as the one adopted here --- there exists no value of $a$ for which it is satisfied by every $\varphi\in D(\underline{\hat{H}}_a)$. In yet other cases, the condition holds for all values of $a$~\cite{Kiefer_2019}. More importantly, the criterion lacks a clear physical justification~\cite{Gotay:1983kvm,Bojowald:2007ky}. As we have seen, imposing $\varphi(0)=0$ is not necessary for the existence of a well-defined quantum dynamics: self-adjointness of $\underline{\hat{H}}_a$ alone guarantees that an evolved state $\varphi(T,v)$ exists for all $T\in(-\infty,\infty)$. By contrast, the classical dynamics ends at a finite time $T=T_{\mathrm c}$, when the singularity is reached and the physical volume vanishes, $V=0$. One may then ask whether the quantum evolution could nevertheless pass through a singular configuration in which the expectation value of the volume operator vanishes, $\big\langle \underline{\hat{V}} \big\rangle\coloneq \expval{\underline{\hat{V}}}{\varphi(T)}=0$. This is, however, ruled out by the unitarity of the quantum evolution~\cite{Gotay:1983kvm}. In this precise sense, the Wheeler–DeWitt quantum dynamics is non-singular, independently of the values of $\varphi(T,0)$ or $a$.

This notwithstanding, even though it has been argued~\cite{Lund} that the probability of the occurrence of a singular state due to $\varphi(0)\neq0$ is negligible compared to that of a non-singular state, allowing $\varphi(0)\neq0$ remains conceptually unsatisfactory. Imposing $\varphi(0)=0$ would formally eliminate this issue. However, it would not prevent the dynamics from reaching states for which $\big\langle \underline{\hat{V}} \big\rangle$ becomes arbitrarily small, which is still physically unreasonable. The reason is that $\varphi(0)=0$ is a purely kinematical restriction, entirely disconnected from the quantum dynamics and the construction of physical observables. Singularity resolution, and more generally any physically meaningful property of the quantum theory, should instead emerge from the dynamics or from the physical observables themselves, rather than being imposed externally as a boundary condition.

In conclusion, there does not seem to exist any consistency requirement capable of fixing the value of the extension parameter $a$. The quantum dynamics is well-defined and non-singular for any choice of $a$. From the perspective of the Wheeler-DeWitt theory, $a$ is therefore a free parameter of the model, whose status is analogous to that of the factor-ordering ambiguities in the operator expression of the Hamiltonian $\underline{\hat{H}}$. Different values of $a$ lead to genuinely different quantum dynamics and, consequently, to distinct physical predictions. The physically correct choice of $a$ must ultimately be determined by experiment.

We can gain further insight into the non-singular dynamics generated by $\underline{\hat{H}}_a$ by studying the evolution of wave packets. Since this problem has already been extensively investigated in the literature (see, for example,~\cite{Gotay:1983kvm,Kiefer_2019}), we will limit ourselves to outlining the main results of these analyses. While different choices of spectral profiles naturally lead to different quantitative details, they all exhibit the same qualitative behavior. If the initial wave packet is suitably peaked on a classical collapsing solution, its initial evolution closely follows the infalling classical trajectory. However, instead of reaching the classical singularity, quantum effects become significant after a finite time, and the expectation value $\big\langle \underline{\hat{V}} \big\rangle$ reaches a nonzero minimum. Subsequently, $\big\langle \underline{\hat{V}} \big\rangle$ increases again, and the wave packet asymptotically follows a classical expanding solution. The resulting quantum dynamics therefore describes a bouncing solution that avoids the classical singularity.

In the precise sense that the evolution is complete and $\big\langle \underline{\hat{V}} \big\rangle$ never vanishes, this quantum dynamics can be regarded as non-singular. However, in the absence of a universal lower bound on $\big\langle \underline{\hat{V}} \big\rangle$, there necessarily exist states whose bounce occurs arbitrarily close to the classical singularity. Indeed, in~\cite{Kiefer_2019}, a wave packet describing the collapse of a dust cloud of solar mass is considered, and it is found that the bounce takes place at an extremely sub-Planckian scale. Since quantum-gravitational corrections are expected to become relevant well before such scales are reached, this represents a serious shortcoming of the Wheeler–DeWitt theory. As we will see in the next subsection, the loop quantum model exhibits the same qualitative bouncing behavior found here. Crucially, however, owing to its different construction of physical observables, the expectation value $\big\langle \hat{\rho}{}_{{}_{\mathrm G}} \big\rangle$ of the loop global density operator $\hat{\rho}{}_{{}_{\mathrm G}}$ is bounded from above. As a consequence, for shells enclosing a total mass-energy $m$ larger than the Planck energy, the bounce necessarily occurs at volumes greater than the Planck volume.

Before moving on, let us make a final comment on the choice of time in models of gravitational collapse. In the physical scenario under consideration, the dust time $T$ is clearly the most natural choice. However, it is by no means unique. While different choices of time are classically equivalent, they generally lead to inequivalent quantum theories. This ambiguity is one aspect of the well-known problem of time in quantum gravity. Following~\cite{Gotay:1983kvm}, in models of gravitational or cosmological collapse whose classical dynamics either begins or ends in a singularity, choices of time $t$ can be divided into two categories. Fast times are those for which the singularity is reached only asymptotically as $t\to\pm\infty$, whereas slow times are those for which the singularity is reached at a finite value of $t$. The dust time $T$ employed here therefore belongs to the latter class. Gotay and Demaret~\cite{Gotay:1983kvm} observed that Wheeler-DeWitt quantum theories constructed using fast times generally remain singular, closely mirroring their classical counterparts, while those based on slow times are non-singular in the precise sense discussed above.

Judging the resulting bouncing dynamics to be excessively exotic, together with several additional considerations, they deemed slow-time quantization to be unphysical. They thus concluded that the quantum theory of gravitational or cosmological collapse should remain singular and ought to be obtained by quantizing the classical theory formulated in a fast time. Nowadays, however, in light of developments in loop quantum gravity, and loop quantum cosmology in particular, where non-singular bouncing dynamics arise naturally and consistently in both slow and fast times, this viewpoint should be reconsidered. The physically correct quantum theory is the non-singular one based on the LQG algebra presented in the next subsection. While Wheeler–DeWitt quantum theories formulated in fast times remain singular, those based on slow times exhibit the same qualitative bouncing dynamics that characterizes the loop theory, albeit with quantitative differences and a non-robust realization of singularity resolution. Wheeler-DeWitt quantum theories formulated in slow times should therefore be preferred over those formulated in fast times, as they appear capable of providing a meaningful glimpse into the loop quantum physics of these models.

\subsection{Loop quantum theory} \label{sec:3.2}

The loop quantization of the symmetry-reduced model consists in constructing a quantum representation of the reduced holonomy-flux algebra via the GNS construction, closely following the strategy employed in full loop quantum gravity. All technical details can be found in~\cite{Ashtekar:2003hd,Ashtekar_2006,Velhinho_2007} and references therein. In the representation where the operator corresponding to the point holonomies in \cref{Nholonomy} is diagonal, the Hilbert space takes the form $L^2(\mathbb{R}_{\mathrm B}, d\mu_{\mathrm B}) $, where $\mathbb{R}_{\mathrm B}$ denotes the Bohr compactification of the real line and $d\mu_{\mathrm B}$ is the Haar measure on it. In this paper, however, we adopt the complementary representation in which the operator corresponding to $v$ is diagonal, yielding $\mathscr{H}^{(x)}=L^2(\mathbb{R}, d\mu_{\mathrm d})$, where $d\mu_{\mathrm d}$ denotes the discrete measure on the real line. An orthonormal basis of $\mathscr{H}^{(x)}$ is provided by the eigenstates $\ket{v}$ of $\hat{v}_x$, satisfying
\begin{equation}
    \braket{v}{v'}=\delta_{v \!\, v'} \, .
\end{equation}
Although the states $\ket{v}$ form an uncountable set, they are nonetheless normalized with respect to the Kronecker delta rather than the Dirac delta. As a consequence, the Hilbert space $\mathscr{H}^{(x)}$ is non-separable.\footnote{While $\mathscr H^{(x)}$ is non-separable, each superselection sector $\mathscr{H}^{(x,\epsilon)}$, dynamically defined and preserved by $\hat{H}$ together with the physical observables, is a separable Hilbert space.} An element $\ket{\psi} \in\mathscr{H}^{(x)}$ can then be expanded as
\begin{equation}
    \ket{\psi} = \sum_ {v\in \mathbb{R}} \psi (v) \ket{v}, \quad\quad \text{with} \quad\quad \sum_{v\in\mathbb{R}} \abs{\psi(v)}^2 < \infty,
\end{equation}
where only a countable subset of values $v\in \mathbb{R}$ contributes non-trivially to the sum. The scalar product in the $v$-representation is therefore given by 
\begin{equation}
    \braket{\psi_1}{\psi_2} =\sum_{v \in \mathbb{R}} \; \psi_1^*(v)  \, \psi_2(v) \, .
\end{equation}
We thus see that the loop quantization departs from the WDW quantization right from the outset.

The classical elementary variables $\mathcal N_\mu(x)$ and $E^x(x)$ are promoted to operators $\big(\widehat{\mathcal N}_{\mu}\big)_x$ and $\big(\hat{E}^x\big)_x$ acting on the Hilbert space $\mathscr{H}^{(x)}$, with commutation relations that are in direct correspondence with the classical holonomy-flux Poisson brackets in \cref{Pbracket2}. Likewise, the variables $\mathcal N(x)$ and $v(x)$ are promoted to operators $\widehat{\mathcal N}_x$ and $\hat{v}_x$ on $\mathscr{H}^{(x)}$ as well, satisfying $\big[\, \widehat{\mathcal N}_x, \hat{v}_x\big]= - \widehat{\mathcal N}_x$ in agreement with \cref{Pbracket3}. For notational simplicity, we suppress the explicit dependence of the operators on the shell label $x$ in what follows. As shown in~\cite{Ashtekar_2006_2}, it is not possible to choose a representation in which both $\widehat{\mathcal N}_\mu$ and $\widehat{\mathcal N}$ act simply. Since a simple action of $\widehat{\mathcal N}$ is necessary to compute the Hamiltonian's action, we have chosen a representation in which $\widehat{\mathcal N}$ has a particularly simple realization. In this $v$-representation, the operator $\hat{v}$ acts multiplicatively, and the action of $\widehat{\mathcal N}$ is uniquely fixed by the above commutation relation to be a unit translation in $v$:
\begin{align*}
    &\hat{v} \ket{v}  = v \ket{v}  \\
    &\hat{\mathcal{N}} \ket{v} = \ket{v+1} \, .
\end{align*}

By studying the matrix elements of $\widehat{\mathcal N}_\mu$, it is straightforward to show that $\widehat{\mathcal N}_\mu$ fails to be weakly continuous in $\mu$ at $\mu=0$~\cite{Ashtekar:2011ni}. This fact has two important consequences. First, the Stone-von Neumann uniqueness theorem does not apply to this representation, since weak continuity is one of its hypotheses. The loop representation turns out in fact to be unitarily inequivalent to the WDW representation discussed in the previous subsection. Second, the lack of weak continuity implies that an operator $\hat{b}$ (or, equivalently, $\hat{K}_\varphi$) cannot 
exist in this representation, since the standard procedure of defining it as the $\mu\to 0$ limit of its exponentiated form, namely the holonomies $\widehat{\mathcal N}_\mu$, is not well defined. As discussed at the beginning of this section, this feature has a clear physical interpretation: it is a direct consequence of the underlying discrete quantum geometry. Following \cref{K_to_sin}, the extrinsic curvature information encoded in $b$ should therefore be expressed in the quantum theory as
\begin{equation} \label{b_to_sin}
    \bar{\mu} K_\varphi =b\longrightarrow \widehat{\sin b}\coloneq\frac{\big(\widehat{\mathcal N}\,\big)^2 - \big(\widehat{\mathcal N}^{-1}\big)^2}{2i} \, .
\end{equation}
In the large $v$ regime, where the granularity of the geometry becomes negligible, this difference operator  approaches its WDW analogue in \cref{b_WDW}~\cite{Ashtekar_2006_2}.

Two observables that will be of particular interest in our analysis are the volume and global density operators $\hat{V}$ and $\hat{\rho}{}_{{}_{\mathrm G}}$, defined as
\begin{align}
    \hat{V} &=2 \pi G \gamma \sqrt{\Delta}  \hbar \,\abs{\hat{v}} ,\\
    \hat{\rho}{}_{{}_{\mathrm G}} &= \rho_c \,\big(\,\widehat{\sin b}\,\big)^2  ,
\end{align}
starting from their classical expressions in \cref{Vv,rhoG}.

With a quantum expression for $b$ at hand, we are now ready to promote the Hamiltonian $H$ in \cref{H_bv} to an operator $\hat{H}$ on $\mathscr{H}^{(x)}$. As in the Wheeler-DeWitt theory, several factor orderings are possible. We adopt a particularly convenient choice, known in the LQC literature as the MMO factor ordering~\cite{MMO}. Although different factor orderings lead to qualitatively similar dynamics, particularly for semiclassical states~\cite{Prescriptions}, they correspond to physically distinct possibilities whose validity must ultimately be settled by experiment. With the chosen prescription, the Hamiltonian operator $\hat{H}$ takes the form
\begin{equation}\label{HLQG}
    \hat{H}=- \frac{3 \hbar}{4 \gamma \sqrt{\Delta}} \abs{\hat{v}}^{\frac{1}{4}} \hat{D}\abs{\hat{v}}^{\frac{1}{2}} \hat{D} \abs{\hat{v}}^{\frac{1}{4}}  ,
\end{equation}
with $\hat{D}\coloneq\frac{1}{2}\left( \sgn(\hat{v}) \, \widehat{\sin b}+\widehat{\sin b} \, \sgn(\hat{v}) \right)$. Its action on a state $\psi(v)$ reads
\begin{equation} \label{H_action}
    \hat{H} \psi(v)= f_+(v) \,\psi(v+4) +f_0(v) \,\psi(v) + f_-(v) \,\psi(v-4) \!\, ,
\end{equation}
where
\begin{align}
    f_+(v)&= \frac{3 \hbar}{64 \gamma \sqrt{\Delta}} s_+(v)s_+(v+2) \abs{v}^{\frac{1}{4}}  \abs{v+2}^{\frac{1}{2}}  \abs{v+4}^{\frac{1}{4}}, \\
    f_0(v)&= -\frac{3 \hbar}{64 \gamma \sqrt{\Delta}}  \abs{v}^{\frac{1}{2}} \left(  s_+(v)^2  \abs{v+2}^{\frac{1}{2}}  +  s_-(v)^2  \abs{v-2}^{\frac{1}{2}} \right) ,\\
    f_-(v)&= \frac{3 \hbar}{64 \gamma \sqrt{\Delta}} s_-(v)s_-(v-2) \abs{v}^{\frac{1}{4}}  \abs{v-2}^{\frac{1}{2}}  \abs{v-4}^{\frac{1}{4}} ,
\end{align}
and $s_\pm(v)\coloneq\sgn(v \pm 2) + \sgn(v)$.

The Hamiltonian $\hat{H}$ is thus represented by a difference operator in the loop quantum theory. Its action exhibits several noteworthy properties:
\begin{itemize}
     \item[i.] The state $\ket{v=0}$ is annihilated by $\hat{H}$ and is dynamically decoupled from the rest of the Hilbert space.

     \item[ii.] $\hat{H}$ commutes with the parity operator $\hat{\Pi}$, which implements a reversal of the triad orientation, $\hat{\Pi}\,\psi(v) = \psi(-v)$.

     \item[iii.] The sectors $v>0$ and $v<0$ are decoupled.

     \item[iv.] $\hat{H}$ couples only points $v\in\mathbb{R}$ separated by steps of 4.
\end{itemize}
While properties i. and iii. are a consequence of the specific factor ordering chosen, the remaining properties hold for any factor ordering, as they follow directly from the classical form of the Hamiltonian. In particular, property ii. is a consequence of the absolute value on $v$ and the quadratic term $b^2$, which together make the Hamiltonian parity invariant regardless of how these terms are split in the quantum theory. Similarly, due to the $b^2$ term, the operators $\mathcal N$ and $\mathcal N^{-1}$ always appear in groups of four, giving rise to the characteristic steps of 4 in property iv.

Property i. implies that the singular state $\ket{v=0}$ is dynamically isolated, so that $v=0$ can be excluded from the support of states $\psi(v)$. This resolves the issue encountered in the Wheeler–DeWitt theory of interpreting the meaning of a hypothetical value $\psi(0)$, since this quantity simply does not belong to the wavefunction in the present framework. Importantly, this exclusion arises directly from the dynamical action of the Hamiltonian itself, rather than being imposed as an external (boundary) condition. However, as already emphasized in the Wheeler–DeWitt theory, this fact alone does not allow one to conclude that the loop theory is non-singular, since the issue of singularity resolution is largely independent of the value or existence of $\psi(0)$.

Properties ii.-iv. imply that the action of the Hamiltonian, together with its symmetry under reversal of the triad orientation, decomposes the Hilbert space $\mathscr{H}^{(x)}$ into dynamically preserved sectors. Moreover, since we are only interested in observables that preserve this decomposition, such as $\hat{V}$ and $\hat{\rho}{}_{{}_{\mathrm G}}$, these sectors define genuine superselection sectors of the theory. Property ii. implies that $\mathscr H^{(x)}$ decomposes into the even and odd eigenspaces of the parity operator $\hat{\Pi}$, while property iii. further allows one to restrict the support of the corresponding states $\psi(v)$ to either $v>0$ or $v<0$. Choosing the sector $v>0$ without loss of generality, property iv. then implies that the Hilbert space further decomposes into sectors spanned by wavefunctions $\psi (v)$ supported on positive semi-lattices of the form
\begin{equation} \label{lattice}
     L_\epsilon \coloneqq \{v=\epsilon+4n, n \in \mathbb{N}, \; \epsilon \in \left(0, 4\right]\} \, .
\end{equation}
The parameter $\epsilon$ labels the lattice and thereby selects a superselection sector $\mathscr{H}^{(x,\epsilon)}$, which inherits its Hilbert space structure from $\mathscr H^{(x)}$. While the full Hilbert space $\mathscr H^{(x)}$ is non-separable, each $\mathscr{H}^{(x,\epsilon)}$ is a separable Hilbert space.

As shown in \cref{B}, and in contrast with the Wheeler-DeWitt case, the operator $\hat{H}$ is essentially self-adjoint. Furthermore, it can be shown~\cite{MMO,Kaminski:2007gm} that its spectrum is purely absolutely continuous and given by $\mathrm{Sp}(\hat{H})=(-\infty,0]$. A basis of $\mathscr{H}^{(x,\epsilon)}$ can then be constructed out of the negative-eigenvalue generalized eigenfunctions $\mathrm e_k(v)$, defined by
\begin{equation}
    \hat{H} \mathrm e_k(v)= -\omega_k^2 \mathrm e_k(v) \quad\quad \text{with} \quad \quad \omega_k^2 =\frac{3\hbar}{4\gamma\sqrt{\Delta}} k^2>0,
\end{equation} 
and satisfying the recursive relation
\begin{equation} \label{ek_recursive}
    \mathrm e_k (v+4)= -\frac{\omega_k^2 + f_0(v)}{f_+(v)} \mathrm e_k(v)- \frac{f_-(v)}{f_+(v)} \mathrm e_k(v-4) \, .
\end{equation}
The arbitrary parameter $k\in{\mathbb R^+}$ labeling these eigenfunctions is chosen in analogy with the Wheeler-DeWitt case in \cref{ek_WDW}. This choice facilitates the comparison between the loop eigenfunctions $\mathrm e_k$ and their WDW counterparts $\underline{\mathrm e}_k$ in the large $v$ regime (see \cref{C}). Since $f_-(\epsilon)=0$ for all $\epsilon\in(0,4]$ --- a property that, together with $f_+(-\epsilon)=0$, is responsible for the decoupling of the sectors $v>0$ and $v<0$ --- the eigenfunction $\mathrm e_k(v)$ within a given superselection sector $\mathscr{H}^{(x,\epsilon)}$ is completely determined by a single initial datum $\mathrm e_k(\epsilon)$. As a consequence, the spectrum of $\hat{H}$ in each superselection sector is non-degenerate. By appropriately fixing the modulus of $\mathrm e_k(\epsilon)$, the (generalized) basis $\ket{\mathrm e_k}$ can be chosen to satisfy the Dirac-delta normalization condition $\braket{\mathrm e_k}{\mathrm e_{k'}}=\delta(k-k')$. The residual freedom in the phase of $\mathrm e_k(\epsilon)$ may then be fixed in the standard way by requiring $\mathrm e_k(\epsilon)\in \mathbb{R}^+$. This choice, together with the reality of the coefficients $f_\pm (v)$ and $f_0 (v)$, guarantees that the $\mathrm e_k(v)$ are real, as is manifest from the recurrence relation in \cref{ek_recursive}. Finally, as shown in \cref{C}, the large-$v$ behavior of the normalized eigenfunctions is
\begin{equation} \label{asymptotic_ek}
    \mathrm e_k (v) = 2 \Big( e^{i\alpha(k)} \underline{\mathrm e}_{k}(v) + e^{-i\alpha(k)} \underline{\mathrm e}^*_{k}(v)\Big) + O \Big(\abs{\underline{\mathrm e}_{k}(v)} v^{-1/2}\Big),
\end{equation}
showing that, for large $v$, namely in the regime where the discreteness of the underlying geometry becomes negligible, the loop eigenfunctions reduce to a simple linear combination of their WDW counterparts.

Since $\hat{H}$ is essentially self-adjoint (see appendix \ref{B}), the Schrödinger equation
\begin{equation} \label{Schr_LQG}
    i \hbar \frac{\partial }{\partial T} \psi (T, v) = \hat{H} \, \psi (T, v) \, 
\end{equation}
generates a unitary time evolution via the operator $\hat{U} (T)\coloneq \exp\big(-iT\hat{H}/\hbar\big)$. Given an arbitrary initial state expanded in the (generalized) eigenbasis of $\hat{H}$,
\begin{equation}
     \psi(T=0, v) = \int_{0}^{+\infty} dk \, \,\tilde{\psi} (k) \, \mathrm e_k (v),
\end{equation}
with spectral profile $\tilde{\psi} (k) $, its evolution under the dynamics generated by $\hat{H}$ is
\begin{equation} \label{physStatesLQG}
    \psi (T, v)= \int_{0}^{+\infty} dk \, \,\tilde{\psi} (k) \, \mathrm e_k (v)
     \,e^{i\omega_k^2 T/\hbar}.
\end{equation}
Exactly as in the Wheeler-DeWitt theory, the self-adjointness of $\hat{H}$, and hence the unitarity of $\hat{U}$, guarantees a complete quantum evolution and rules out the possibility that the expectation value $\big\langle\hat{V}\big\rangle$ of the volume operator vanishes at any finite time. In this precise sense, the quantum dynamics is non-singular. However, this fact alone does not exclude physically unreasonable scenarios, such as the evolution toward states for which $\big\langle\hat{V}\big\rangle$ becomes arbitrarily small. This was indeed the case in the Wheeler-DeWitt theory.

In the loop quantum theory, by contrast, it is straightforward to show that the global density operator $\hat{\rho}{}_{{}_{\mathrm G}}$ is a bounded self-adjoint operator with spectrum $\mathrm{Sp} \big(\hat{\rho}{}_{{}_{\mathrm G}}\big)=[0,\rho_{\mathrm c}]$. Consequently, the global mass–energy density enclosed by the shell $x$ is universally bounded above by the critical density $\rho_{\mathrm c} = 3/(8\pi G \gamma^2 \Delta)$. We have therefore established that $\rho_{{}_{\mathrm G}}$, which diverges as the shell approaches the singularity in the classical theory, remains finite throughout the quantum evolution in the loop quantum theory. This means that the quantum dynamics cannot access configurations in which this observable attains values arbitrarily close to the ones it assumes at the classical singularity. This yields a notion of singularity resolution that is significantly more robust than the mere completeness of the quantum evolution.

Interestingly, although the volume operator $\hat{V}$ effectively assumes only non-vanishing discrete values within each dynamically induced superselection sector $\mathscr H^{(x,\epsilon)}$, it does not have a universal positive lower bound on $\mathscr H^{(x)}$. The physically correct choice of $\epsilon$, and thus the physically realizable spectrum of $\hat{V}$, cannot be fixed by internal consistency. It should instead be determined either experimentally or, possibly, through a genuine quantum-level symmetry reduction from full loop quantum gravity. Nevertheless, numerical analysis shows that for semiclassical states peaked at a given total mass-energy $m_0$ exceeding the Planck mass, the existence of a global upper bound on the energy density implies a corresponding lower bound on $\big\langle\hat{V}\big\rangle$. During the evolution, this expectation value remains larger than the Planck volume.

To gain deeper insight into the mechanism of singularity resolution in the theory, particularly for initially semiclassical collapsing states, a more detailed analysis of the quantum dynamics is necessary. Although closed analytical expressions for the eigenfunctions $\mathrm e_k(v)$ are available, they are not sufficiently manageable to allow for a comprehensive analytical treatment of the dynamics. We therefore turn to a numerical investigation of the evolution of wave packets.

\subsubsection{Numerical analysis}

We are primarily interested in the evolution of (initially) semiclassical states describing collapsing configurations. By semiclassical state, we mean a state sharply peaked on a classical trajectory. Namely, a state that is sharply peaked around the mean values of the relevant observables, and whose evolution of such mean values approximately follows the classical equations of motion. To this end, we consider wave packets $\psi(T,v)$ of the form given in \cref{physStatesLQG}, with a spectral profile $\tilde{\psi}(k)$ chosen such that the initial state $\psi(T=0,v)$ is peaked around a prescribed volume $V_0 = 2 \pi G \gamma \sqrt{\Delta}  \hbar \, v_0$, conveniently parametrized by $v_0$, and a prescribed total mass-energy $m_0= V_0 \,\rho_0$, where $\rho_0$ denotes the classical global density associated with $V_0$ and $m_0$. For any given set of initial data $(v_0, \rho_0, \tilde{\psi}(k))$, the state $\psi(T,v)$ can then be computed numerically. We briefly outline the procedure below; detailed descriptions of the numerical methods can be found, for example, in~\cite{Prescriptions,Ashtekar_2006_2}.

The first step in the numerical computation of a state $\psi(T,v)$ of the form given in \cref{physStatesLQG} is the numerical evaluation of the loop eigenfunctions $\mathrm e_k (v)$. Although the initial value $\mathrm e_k(\epsilon)\in\mathbb{R}^+$ generating Dirac-delta normalized eigenfunctions is not easily determined analytically, it can be obtained numerically by analyzing the large-$v$ behavior of $\mathrm e_k (v)$ as described in \cref{C}. Since the convergence rate of \cref{asymptotic_ek} is relatively slow, the actual numerical implementation relies on the improved asymptotic relation
\begin{equation} \label{asymptotic_ek_2}
    \mathrm e_k (v) = 2 \Big( e^{i\alpha(k)} \underline{\mathrm e}^{(4)}_{k}(v) + e^{-i\alpha(k)} \underline{\mathrm e}^{(4)}_{k}{}^*(v)\Big) + O \Big(\abs{\underline{\mathrm e}^{(4)}_{k}(v)} v^{-3}\Big),
\end{equation}
where the modified WDW functions $\underline{\mathrm e}^{(4)}_k (v)$ are defined by
\begin{equation}
    \underline{\mathrm e}^{(4)}_k = \frac{\zeta(v)}{\sqrt{2 \pi}}  e^{i k \xi (v)  }  ,
\end{equation}
with
\begin{equation}
    \zeta(v) = \frac{1}{v^{1/4}} + \frac{B_1}{v^{5/4}} +\frac{B_2}{v^{9/4}},
\end{equation}
\begin{equation}
    \xi(v) = v^{1/2} + \frac{A_1}{v^{1/2}}  + \frac{A_2}{v^{3/2}}  + \frac{A_3}{v^{5/2}},
\end{equation}
and the coefficients $A_j$ and $B_j$ are given in \cref{parameters_AB}. Further details are provided in \cref{C}.

The spectral profiles chosen for the wave packets are
\begin{equation} \label{spec_profile}
    \tilde{\psi}(k)=N \exp \left[- \frac{(k-k_0)^2}{2 \sigma^2}-i\biggl( k \, \xi(v_0) + \alpha(k) \biggr) \right] \,
\end{equation}
where $k_0^2=\frac{4 \gamma \sqrt{\Delta}}{3 \hbar}m_0$, $\alpha(k)$ is the phase shift in the asymptotic relation in \cref{asymptotic_ek_2}, $N$ is a normalization constant, and $\sigma$ controls the dispersion of the wave packet. 
For $v\gg 1$, we obtain
\begin{equation} \label{State_early}
    \begin{split}
        \abs{\psi(T, v \gg 1)}^2 & = \abs{ \int_\mathbb{R} dk \; \tilde{\psi}(k)  \mathrm e_k(v) e^{i \omega_k^2 T/\hbar} }^2 \\
        &\approx\abs{ \int_\mathbb{R} dk \; \tilde{\psi}(k) \frac{4\zeta(v)}{\sqrt{2 \pi}} \cos \biggl( k \, \xi(v) + \alpha(k) \biggl) e^{i \omega_k^2 T/\hbar}}^2  \\
        & \approx \,N'(T) \,\zeta(v)\, \exp \left[ - \frac{\left( \xi(v)-\xi(v_0)+\frac{3 }{2 \gamma \sqrt{\Delta} } k_0 T\right)^2}{\frac{1}{\sigma^2}+\frac{9 \sigma^2}{4 \gamma^2 \Delta} T^2} \right],
    \end{split}
\end{equation}
which provides a good approximation to $\abs{\psi(T,v)}^2$ when $v_0 \gg 1$ --- that is, when the peak of the initial state lies in the large-$v$ regime --- and for sufficiently small $T$. The specific phase chosen in \cref{spec_profile} ensures that $\psi(0, v)$ is also approximately Gaussian, provided $v_0$ is sufficiently large. In particular, the function $\xi(v)$ appearing in \cref{spec_profile} is adapted to the improved large-$v$ behavior of $ \mathrm e_k (v)$ given in \cref{asymptotic_ek_2}. This choice extends the range of $v_0$ for which \cref{State_early} accurately approximates $\abs{\psi(T,v)}^2$ and allows the approximation to remain valid over longer evolution times $T$. One could alternatively consider a spectral profile of the form in \cref{spec_profile} with the replacement $\xi(v) \rightarrow \sqrt{v}$. Although an expression analogous to \cref{State_early} could still be derived using the weaker asymptotic relation \cref{asymptotic_ek}, it would provide a good approximation to $\abs{\psi(T,v)}^2$ only for significantly larger values of $v_0$ and for shorter times $T$.

\begin{figure}[b]
    \captionsetup{justification=raggedright}
    \centering
    \includegraphics[width=.55\textwidth]{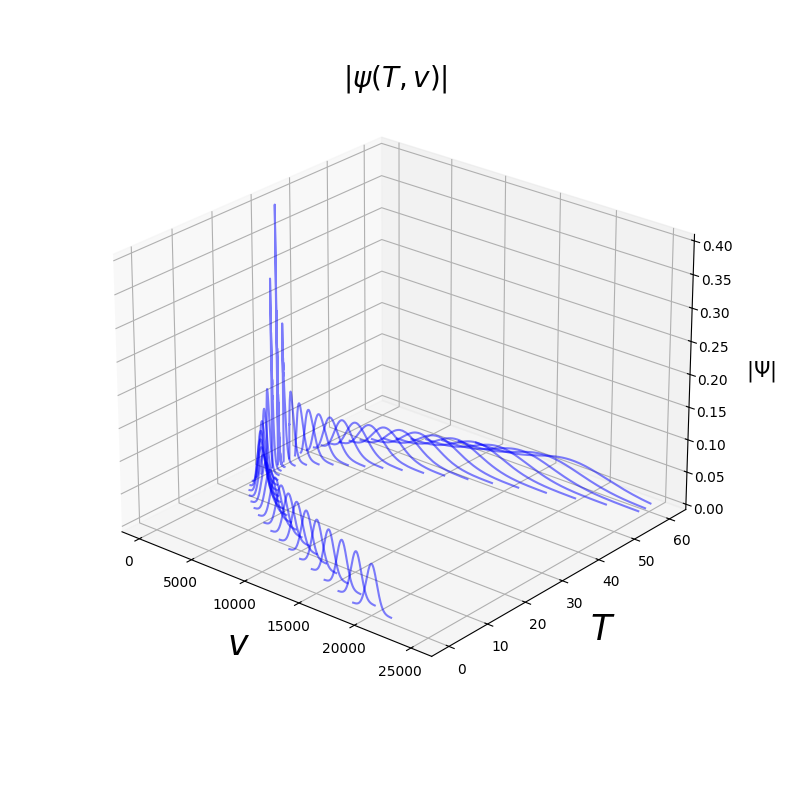}\hfill
    \raisebox{1cm}{
    \includegraphics[width=.4\textwidth]{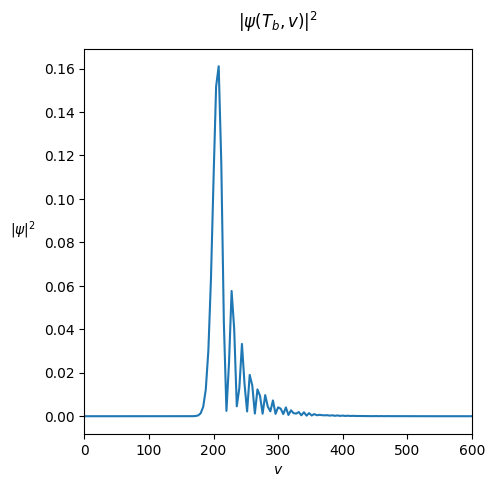}
    }
    \caption{Plots portraying the evolution of $\psi(T, v)$. On the left, a 3D plot showing $\abs{\psi(T, v)}$. On the right, a snapshot of $\abs{\psi}^2$ at the bouncing time $T_b$. Units have been chosen to be $c=\hbar=G=\gamma=1$ and the initial values have been set to $v_0=20000$, $\rho_0=10^{-2} \rho_c$, and $\sigma=0.6$.}
    \label{fig:wavepacket}
\end{figure}

\begin{figure}[ht]
\captionsetup{justification=raggedright}
\centering
\includegraphics[width=.99\textwidth]{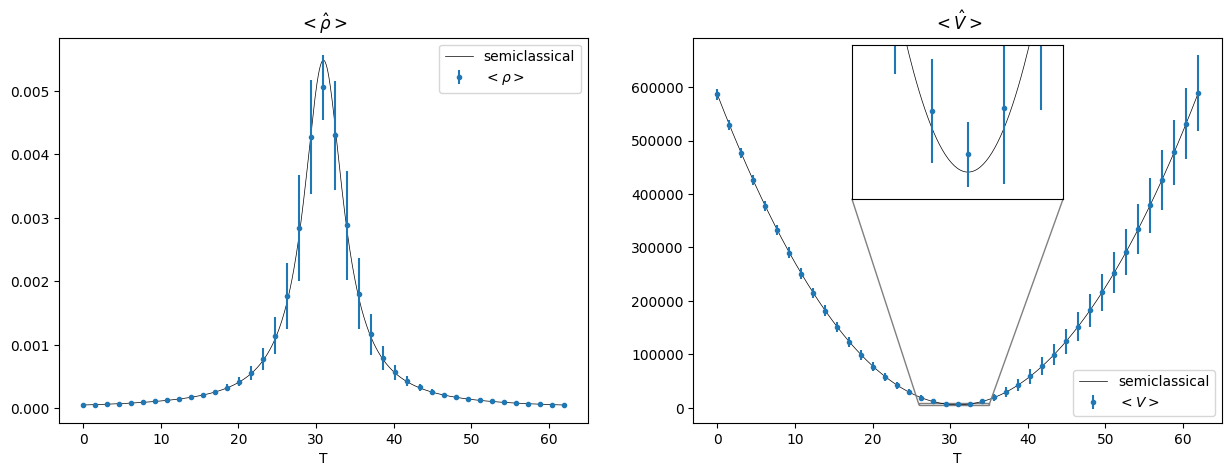}

\caption{Expectation values of $\hat{\rho}{}_{{}_{\mathrm G}}$ and $\hat{V}$ with their variance for the state $\psi(T, v)$ in \cref{fig:wavepacket} compared with the trajectories predicted by the effective semiclassical theory.}
\label{fig:ExpVal}

\end{figure}

The evolution of an initially collapsing wave packet with $v_0\gg 1$ is shown in \cref{fig:wavepacket}, while the corresponding evolution of $\big\langle \hat{\rho}{}_{{}_{\mathrm G}} \big\rangle$ and $\big\langle \hat{V} \big\rangle$ is displayed in \cref{fig:ExpVal}.
%
%
%
%
%
%
%
%
%
%
Four key features of the dynamics exhibited in these figures deserve comment:
\begin{enumerate}

    \item The expectation value $\big\langle \hat{V} \big\rangle$ of the volume enclosed by the shell --- and thus the peak of the state $\psi(T, v)$, which for a sufficiently narrow wave packet is numerically close to $\langle \hat{v} \rangle$ --- initially decreases, reaches a strictly positive minimum at some time $T_b$, and subsequently grows back to large values. The evolution therefore describes a bouncing solution that avoids the classical singularity. In contrast to the bouncing solutions of the WDW theory, the LQG bounce always occurs when $\big\langle \hat{\rho}{}_{{}_{\mathrm G}} \big\rangle$ reaches the Planck scale, thereby providing a precise and physically consistent scale for the onset of quantum effects. This behavior mirrors that of the cosmological universe state in loop quantum cosmology.

    \item Near the bouncing point, the state develops oscillations. To the authors' knowledge, this feature has not been observed in previous numerical investigations within LQC. We return to this point below.

    \item The width of the wave packet, closely related to the variance of the volume operator, generally increases over time, except in the immediate vicinity of the bounce, where oscillations develop. Given two times $T_1 < T_b$ and $T_2 > T_b$ such that the peak of the wave packet satisfies $v(T_1)=v(T_2)$, the width of the state at $T_2$ is larger than at $T_1$. This behavior is however expected. In contrast to LQC with a massless scalar field, where the evolution equation is of the Klein–Gordon type, the dynamics here is governed by a Schrödinger equation. As is also evident from \cref{State_early}, the spreading of the state is entirely analogous to that of a free Gaussian wave packet in ordinary quantum mechanics. By contrast, the variance of the global density exhibits a different behavior: it increases during the bounce and subsequently decreases again in the expanding phase.

    \item From \cref{fig:ExpVal}, we observe that the expectation values agree overall quite well with the effective semiclassical theory (see \cref{D}). However, within a small interval centered around $T_b$, $\big\langle \hat{\rho}{}_{{}_{\mathrm G}} \big\rangle$ and $\big\langle \hat{V} \big\rangle$ are respectively slightly smaller and slightly bigger than their effective counterparts. This deviation can be attributed to the development of oscillations near the bounce. As a consequence, the accuracy of the effective dynamics in reproducing the full quantum evolution in the vicinity of the bounce is sensitive to the formation of these oscillations.
    
\end{enumerate}


The interference pattern shown in the right panel of \cref{fig:wavepacket} is qualitatively similar to that arising in the reflection of a free particle off an infinite potential wall in non-relativistic quantum mechanics. Since our initially collapsing wave packet undergoes a form of reflection as well, this resemblance is perhaps not unexpected. A similar phenomenon can also be observed in the Wheeler-DeWitt quantum theory, as shown in~\cite{Kiefer_2019}. In that case, by comparing \cref{H_WDW2} with the non-relativistic Schrödinger equation, the bounce may be interpreted as a reflection off an effective potential. An analogous interpretation applies in the loop quantum theory, although the evolution equation is now a discrete difference equation rather than a differential one.

More surprising, however, is that, to the authors’ knowledge, such behavior has not been reported in previous studies of the bounce of cosmological wave packets in loop quantum cosmology. Since one of the main differences from most investigations in LQC is the presence of a Schrödinger-type evolution equation, which leads to an unavoidable spreading of the wave packet over time, the observed difference in behavior could be a consequence of a dependence of the interference pattern at the bounce on the width of the wave packet.

\begin{figure}[b]
\captionsetup{justification=raggedright}

\centering
\includegraphics[width=.3\textwidth]{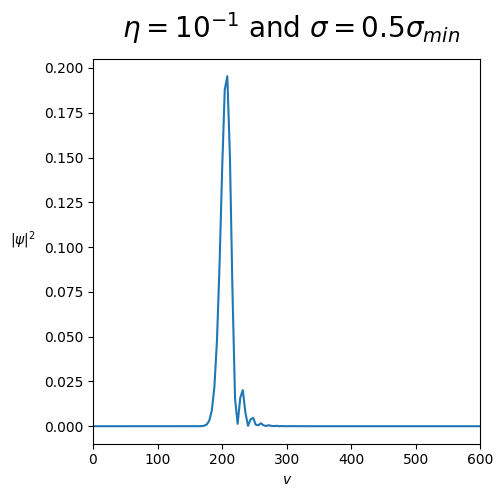}\hfill
\includegraphics[width=.3\textwidth]{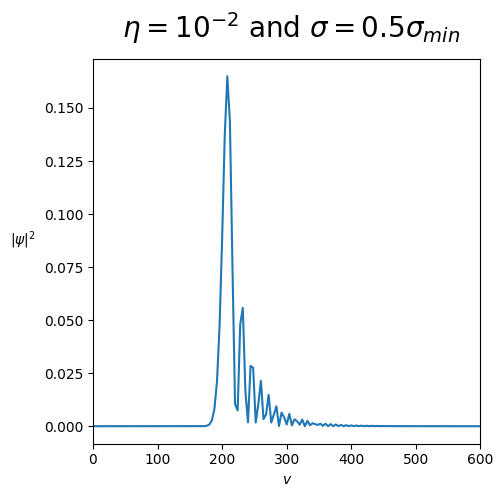}\hfill
\includegraphics[width=.3\textwidth]{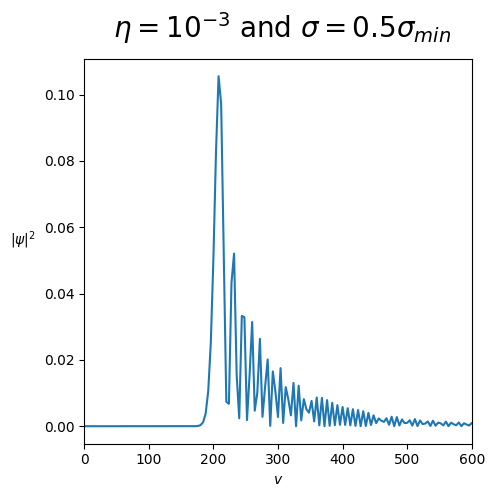}

\vspace{0.5cm}
\includegraphics[width=.33\textwidth]{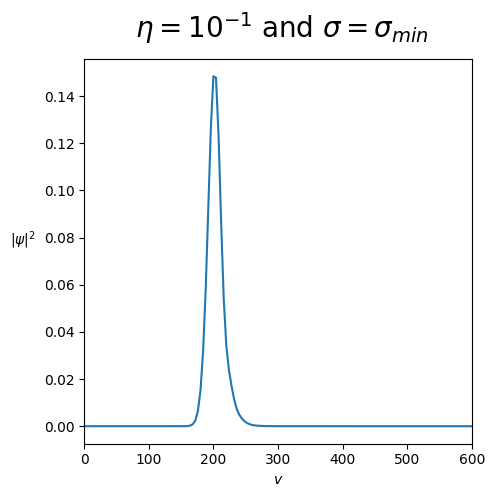}\hfill
\includegraphics[width=.33\textwidth]{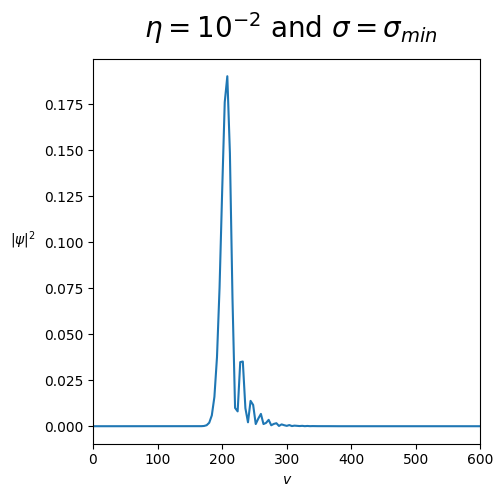}\hfill
\includegraphics[width=.33\textwidth]{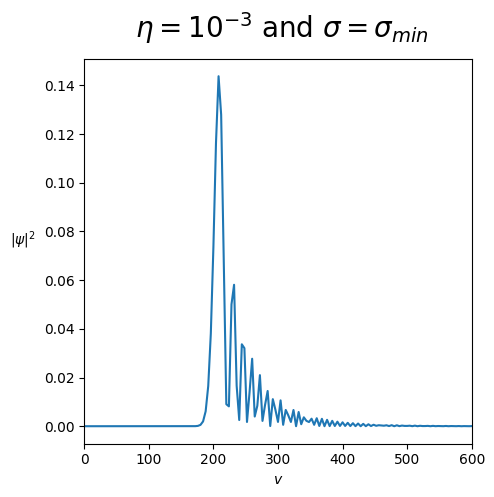}

\vspace{0.5cm}
\includegraphics[width=.3\textwidth]{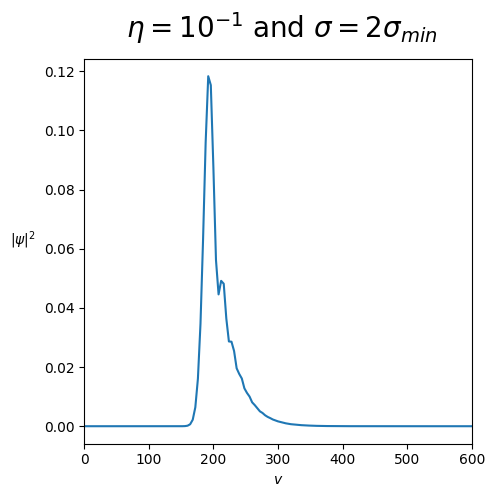}\hfill
\includegraphics[width=.3\textwidth]{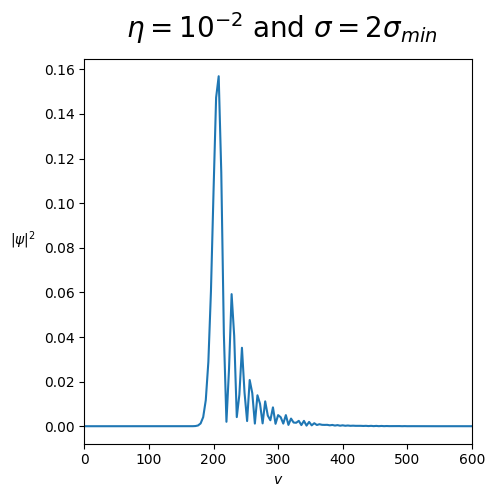}\hfill
\includegraphics[width=.3\textwidth]{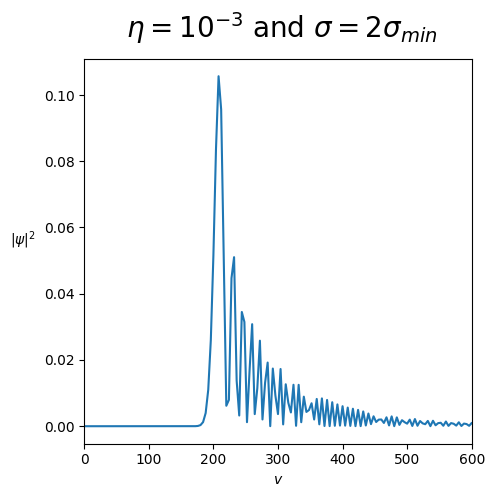}

\caption{A 3x3 table of plots showing $\abs{\psi(T_b, v)}^2$ for different values of $\eta$ and $\sigma$. The initial parameter $v_0$ has been set in such a way that the bouncing volume is the same ($v_b \approx 200$) for all the plots.}
\label{fig:table}

\end{figure}

Let us test this conjecture numerically. The width of the wave packet at the bounce is controlled by two parameters: the bouncing time $T_b$ and the dispersion parameter $\sigma$ in \cref{spec_profile}. The former can be estimated analytically within the effective semiclassical theory (see \cref{D}), yielding
\begin{equation}
    T_b=\frac{2\gamma \sqrt{\Delta}}{3} \sqrt{\frac{1}{\eta}-1},
\end{equation}
where $\eta = \rho_0/\rho_c$. Numerically, one can verify that this expression accurately predicts the time at which the wave packet's peak reaches its minimal value. To reduce the spreading at the bounce, one may therefore decrease the bouncing time by increasing the initial energy density. At early times, when \cref{State_early} provides a reliable approximation to $\abs{\psi(T,v)}^2$, the width of the wave packet is controlled by
\begin{equation} \label{spread}
    \Sigma^2 \coloneqq  \frac{1}{\sigma^2} + \frac{9 \sigma^2}{4 \gamma^2 \Delta} T^2 \, .
\end{equation}
Even though the validity of this expression breaks down before the bounce is reached, one may nevertheless determine the value $\sigma_{\mathrm{min}}$ that minimizes $\Sigma$ at $T_b$:
\begin{equation}
    \sigma_{\mathrm{min}} = \sqrt{\frac{2 \gamma \sqrt{\Delta}}{3 T_b}}=\left( \frac{\eta}{1-\eta}\right)^{1/4}.
\end{equation}
Numerical analysis confirms that this choice indeed minimizes the width of the wave packet at the bounce. As a side remark, although this value optimizes the width at the bounce, it may yield an initial wave packet that is either excessively narrow or overly spread. This issue, however, is not relevant for the present investigation.

\begin{figure}[t]
\captionsetup{justification=raggedright}
\centering
\includegraphics[width=.33\textwidth]{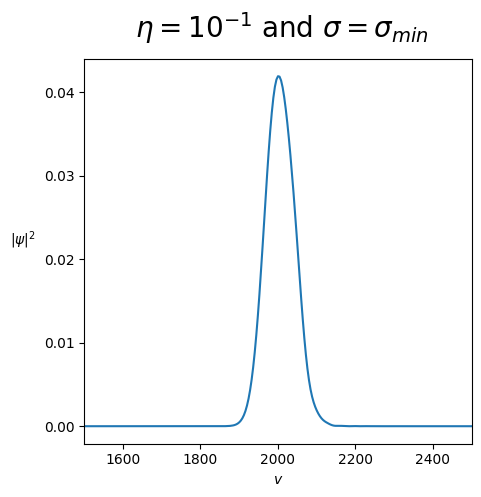}\hfill
\includegraphics[width=.33\textwidth]{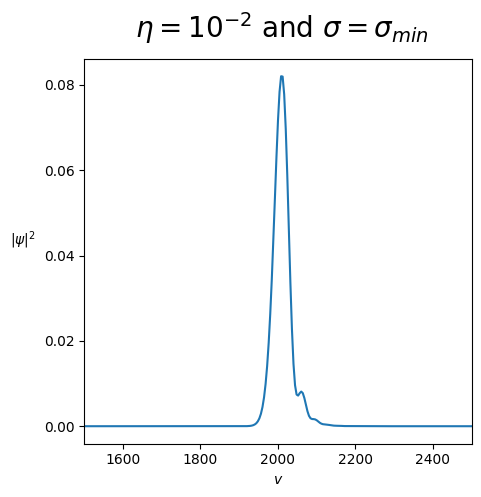}\hfill
\includegraphics[width=.33\textwidth]{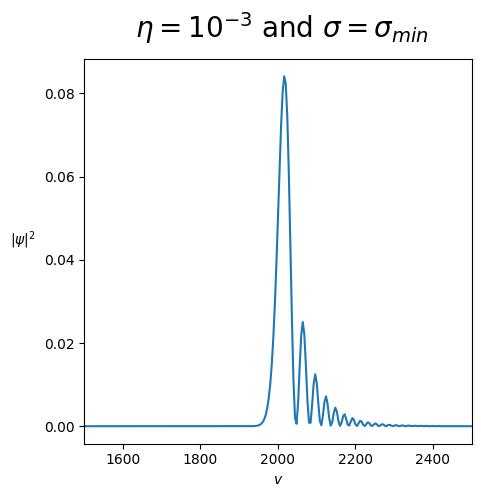}

\caption{Probability density at the bounce $\abs{\psi(T_b, v)}^2$ for three different values of $\eta$ and $\sigma=\sigma_{\mathrm{min}}$. Differently from the configurations shown in \cref{fig:table}, the volume at which the bounce occurs has been increased to $v_b \approx 2000$.}
\label{fig:LargeVb}

\end{figure}

\Cref{fig:table} displays $\abs{\psi(T_b, v)}^2$ for different values of $\eta$ and $\sigma$, arranged in three rows (constant $\sigma$) and three columns (constant $\eta$). To facilitate comparison, we choose $v_0$ such that the bounce takes place approximately at the same value $v_b$ for all three values of $\eta$. From each row, we observe that the oscillations are reduced as $\eta$ increases. Since a larger $\eta$ corresponds to a shorter bounce time $T_b$, this indicates that the interference pattern is reduced when the wave packet has less time to spread. For fixed $\eta$, on the other hand, the oscillations are minimized when $\sigma$ is closest to its optimal value $\sigma_{\mathrm{min}}$. These results clearly demonstrate a strong dependence of the interference pattern on the width of the wave packet at the bounce. In particular, for states that are sufficiently sharply peaked at that time, the interference can even be completely suppressed. Furthermore, as shown by comparing the second row of \cref{fig:table} with \cref{fig:LargeVb}, all else being equal, the interference is less pronounced for wave packets that bounce at larger volumes $v_b$.

This provides a natural explanation for the absence of the interference pattern in previous investigations within loop quantum cosmology. First, when the evolution is governed by a Klein–Gordon-type equation, the wave packet does not undergo dispersion over time, making it easier to reach the bounce with a sharply peaked profile. Moreover, by choosing initial conditions that lead to a bounce at sufficiently large volumes $v_b$, the interference effects can be further suppressed.

Finally, returning to the accuracy of the effective semiclassical theory, the dependence of the formation of oscillations on the bounce volume $v_b$ implies that the accuracy of the effective description itself depends on the position of the shell within the dust cloud. Shells near the boundary are expected to bounce at larger volumes, where interference effects are suppressed and the effective dynamics provide an accurate approximation to the full quantum evolution. By contrast, shells closer to the center necessarily bounce at smaller volumes, where interference effects become significant and the quantum dynamics can deviate appreciably from the effective description. Differently from the spatially flat FLRW cosmological scenario, where the physical scale of the problem is controlled by an arbitrary parameter $V_0$ --- the volume of the so-called ``fiducial cell'' --- which can be rescaled at will to improve the accuracy of the effective equations~\cite{Rovelli:2013zaa}, here the physical scale is determined dynamically by the system itself and cannot be adjusted externally. As a result, the regime of validity of the effective description is no longer universal, but instead depends intrinsically on the physical properties of each shell.

\subsection{Multi-shell quantum theory} \label{sec:3.3}

We have thoroughly analyzed the loop quantum theory of a single shell in the marginally bound LTB model. A proper quantum LTB model should be obtained by quantizing the classical theory described in \cref{sec:2}, without imposing the LTB condition in \cref{LTBcondition}, along the lines of~\cite{Bojowald_2004,Bojowald:2005cb,Gambini_2014,Gambini:2020nsf} and the quantum framework constructed in~\cite{Husain:2021ojz,Husain_2022}. However, the functional character of such a theory makes a detailed understanding of its physical implications considerably more challenging. A natural question is therefore whether the single-shell model can serve as the basis for a quantum theory that captures the qualitative features of the full quantum LTB dynamics. Classically, the evolution of a dust shell in the LTB model is completely independent of that of the other shells in the cloud, except for the occurrence of shell-crossing singularities. Although this independence need not persist at the quantum level, we may adopt it as a working assumption and use it as a simplifying hypothesis to probe the qualitative physics of the full quantum LTB model.

Before constructing this quantum model, it is worth emphasizing that the classical theory discussed in \cref{sec:2} can also be used to describe the vacuum region outside the collapsing matter. However, it is unclear whether the quantization of this vacuum region can be carried out in the same manner as for the matter sector. For this reason, we restrict our attention to the matter region. We then extend the assumption of independence between different shells to encompass both matter shells and ``vacuum shells'', while explicitly constructing the quantum theory only for the former.

We must therefore construct a quantum theory for a collection of non-interacting shells labeled by $x$, with $x \in [0, \tilde{x}]$. A first, natural assumption is:
\begin{itemize}
    \item[A1)] The dynamical discretization which occurs on the spectrum of each operator $\hat{v}_x$ is identical for all $x \in[0, \tilde{x}]$. In simple terms, this fixes the $\epsilon$ parameter to be the same for every shell, implying that their dynamics happens on the same lattice $L_\epsilon$. 
\end{itemize}
However, the very notion of a shell may depend on the context. So far, shells have been treated as infinitesimally thin two-spheres with no volume, resulting in an uncountable family indexed by $x \in [0, \tilde{x}]$. While this standard idealization is particularly convenient in the classical theory, it poses significant challenges in the multi-shell quantum setting we want to develop, as it would naturally lead to an infinite tensor-product structure for the Hilbert space. To circumvent this difficulty, one may discretize the coordinate $x$ and consider a finite collection of two-spheres arranged on a lattice with arbitrary step $\delta x$. From a more physically motivated perspective, one may instead abandon the idealization of infinitesimally thin shells and consider shells with finite thickness parametrized by $\delta x$. This again results in a finite number of shells separated by an arbitrary step $\delta x$, effectively yielding the same lattice configuration in $x$.

Using the initial volume $V(x)$ rather than the coordinate $x$ itself to label the shells, a natural choice for the lattice is the one that already arises in the quantum theory of a single shell from the discretization of the volume variable. This leads to our second assumption:
\begin{itemize}
    \item[A2)] We take the lattice effectively discretizing the classical theory to be $L_\epsilon$ (see \cref{lattice}). In this picture, the dust star is classically described as a collection of $N$ shells, each associated with a lattice point $v_i=\epsilon +4 i$, representing the increasing volume $V_i$ enclosed by the $i$-th shell, up to a maximal value $v_N$ corresponding to the initial total volume $V_N$ of the star. 
\end{itemize}
It is worth emphasizing that this classical discretization and the discrete spectrum of the volume operator in the quantum theory of a single shell are conceptually independent. The former serves merely as a regulator of the classical continuum description, whereas the latter reflects the intrinsic discreteness of the underlying quantum geometry. The fact that both structures are described by the same lattice $L_\epsilon$ is therefore a simplifying assumption rather than a fundamental requirement.

With this assumption, the multi-shell quantum theory can be constructed in a straightforward manner starting from the single-shell theory discussed in the previous section. The Hilbert space $\mathscr{H}$ of the theory is given by the tensor product 
\begin{equation}
    \mathscr{H}= \bigotimes\limits_{i=1}^{N} \mathscr{H}^{(x_i,\epsilon)}\,.
\end{equation}
Formalizing the assumption of non-interaction between shells as
\begin{itemize}
    \item[A3)]  The system consists of $N$ non-interacting and distinguishable shells,
\end{itemize}
the physical states $\Psi$ of the system must be separable. Namely, they can be written as a tensor product of the single-shell states $\psi_i \in \mathscr H^{(x_i,\epsilon)}$: 
    \begin{equation} \label{total_state}
        \Psi(v_1, v_2 \, , ... \, , v_N)=\psi_{1}(v_1) \,\psi_{2}(v_2) \cdot\, ... \, \cdot \,\psi_{N}(v_N) \, .
    \end{equation}
Single-shell operators are naturally extended to $\mathscr{H}$. Let $X$ denote any operator discussed in \cref{sec:3.2}. We can then define $\hat{X}_i \coloneqq \mathbbm{1}\otimes \, ... \, \otimes \hat{X}_{x_i} \otimes \, ... \, \otimes \mathbbm{1} $, whose expectation value reads
\begin{equation}
   \big\langle\hat{X}_i \big\rangle= \expval{\hat{X}_{x_i}}{\psi_{i}} \, .
\end{equation}
Due to assumption (A3), the Hamiltonian operator $\hat{H}$ of the $N$-shell system is simply $ \hat{H}=\sum_i \hat{H}_i$, so that the time-evolved state is given by
\begin{equation}
    \Psi(T,v_1, v_2 \, , ... \, , v_N)=\psi_{1}(T,v_1) \,\psi_{2}(T,v_2) \cdot\, ... \, \cdot \,\psi_{N}(T,v_N) \, .
\end{equation}
In other words, the dynamics of the full system is entirely determined by the single-shell evolution.

In the classical theory, the evolution generically leads to the formation of shell-crossing singularities~\cite{SCS}. These occur when the trajectories of two neighboring shells intersect, producing a physical, albeit weak, curvature singularity at which the local energy density (see \cref{EOMLTB})
\begin{equation} \label{rho_local}
\rho = \frac{1}{G}\frac{\partial_x m}{\partial_x V}
\end{equation}
diverges. Although the spacetime can be extended beyond such events~\cite{Newman_1986}, the assumption of non-interaction between shells, which underlies the physical reliability of the model, breaks down. In recent years, the formation of shell-crossing singularities has been extensively investigated in the semiclassical framework of LQG effective models~\cite{Kelly_2020_2,Husain_2022,Fazzini:2023scu,Fazzini_2024,GieselGeneralized,Giesel:2024mps,Bobula:2024chr,Cafaro}. The multi-shell model constructed above offers the possibility of going beyond the semiclassical approximation and analyzing the formation of shell-crossing singularities within a fully quantum setting. Motivated by \cref{rho_local}, one may introduce a notion of local energy density operator for the $i$-th shell ($i\neq 1$) in the quantum theory, 
\begin{equation} \label{rho_local_quantum}
    \hat{\rho}^{\mathrm{local}}_i \coloneq \frac{1}{G}\frac{\hat{H}_{i-1}-\hat{H_{i}}}{\hat{V}_i-\hat{V}_{i-1}}.
\end{equation}
The expectation value of this operator can then be investigated numerically for different initial multi-shell states in order to study the development of shell-crossing singularities in the quantum theory as determined by the single-shell dynamics.

The model has clear limitations. It is not obtained through a proper canonical quantization of the full LTB model, but rather constructed by combining multiple single-shell models. As such, it neglects possible interactions of genuinely quantum origin between shells that have no counterpart in the classical theory. Such interactions could play a central role in the overall evolution of the multi-shell state and, in particular, in the formation of shell-crossing singularities. Nevertheless, since the proper quantum theory is currently too complex to allow for a detailed dynamical analysis, the multi-shell model provides a valuable framework in which to investigate the evolution of multiple shells within a fully quantum setting. Furthermore, should interaction terms between shells be identified in the Hamiltonian of the proper quantum theory, they could be systematically incorporated into the multi-shell Hamiltonian and studied in isolation within this simplified framework.

An additional open question is whether this model can be used to understand the quantum nature of the trapped region and horizon, which are bound to form classically. In particular, it would be worthwhile to explore whether a fully quantum treatment could explain the interplay between a (quantum) trapped region in the pre-bounce phase and an anti-trapped one in the post-bounce phase, as suggested in~\cite{Han:2023wxg}. Such an analysis could also shed light on the global structure of the semiclassical limit of spacetime, including geodesic completeness, the nature of the horizon(s) and, closely related to the latter, presence of mass inflation~\cite{Hamilton:2008zz} and its stability properties.

A comprehensive analysis of the model and its full range of implications requires considerably more work and is therefore left for future investigation.

\section{Summary and Conclusions} 
\label{sec:4}

In this paper, we have presented a loop quantization of the marginally bound LTB model, which describes the gravitational collapse of pressureless dust, by constructing the quantum theory of the full system starting from that of a single shell. This result is made possible by the reduction procedure discussed in \cref{sec:2}, where a constraint known as the LTB condition is used to obtain a single-shell Hamiltonian that is free of radial derivatives. Under the assumption that the classical non-interaction among shells carries over to the quantum regime, the quantum LTB model can be treated as a collection of $N$ non-interacting shells, each governed by the single-shell loop quantum theory presented in \cref{sec:3.2}. This model provides an ideal arena in which the evolution of multiple shells can be investigated in a fully quantum gravitational setting, including the possibility of probing the formation of shell-crossing singularities beyond the semiclassical approximation.

We have shown that the single-shell loop quantum theory is non-singular. In close analogy with the behavior of universe states in loop quantum cosmology, a wave packet initially peaked on a collapsing classical trajectory undergoes a quantum bounce when the density of total mass-energy enclosed within the shell approaches the Planckian scale, subsequently following a classical expanding trajectory. The classical central curvature singularity is therefore resolved in both the single-shell and multi-shell quantum theories.

Although the Wheeler-DeWitt quantum theory discussed in \cref{sec:3.1} is likewise non-singular and displays a qualitatively similar physics --- featuring a complete quantum evolution and a bouncing dynamics for initially collapsing wave packets --- it lacks the loop bound on mass-energy density, resulting in a non-robust realization of singularity resolution. Interestingly, whereas the loop quantum theory remains non-singular regardless of the choice of time, the Wheeler-DeWitt quantum theory is non-singular only for slow times, such as the dust time $T$ adopted here, and becomes singular for fast times~\cite{Gotay:1983kvm}. Among Wheeler-DeWitt quantum theories, those formulated in slow times ought therefore to be preferred over those formulated in fast times, given their apparent ability to reproduce the key aspects of the loop quantum physics of these models.

The numerical investigation of the loop quantum dynamics has revealed that initially collapsing wave packets generically develop an interference pattern at the bounce. This behavior, which is a well-known feature of wave packet scattering in non-relativistic quantum mechanics, can be understood as arising from the reflection of the wave packet off an effective potential as dictated by the evolution dynamics. An in-depth analysis of this phenomenon indicates that its effects are more pronounced for broader wave packets and when the bounce occurs at smaller volumes. This accounts for why this phenomenon has not been previously reported in the context of loop quantum cosmology, where typical initial conditions are such that the interference is entirely suppressed.
As a result of this interference, the effective LQG theory no longer provides an accurate description of the quantum dynamics of shells near the center of the dust cloud, where volumes are small. Nevertheless, it continues to faithfully capture the dynamics of the outermost shells, for which the interference at the bounce remains negligible.

We further observed that the spreading of the wave packet causes the post-bounce phase to appear more quantum than the pre-bounce one. This is an unavoidable feature of the Schrödinger equation and is directly tied to the choice of matter content of the model. In LQC, coupling gravity to a massless scalar field or radiation yields a Klein-Gordon type evolution equation, which does not give rise to wave packet spreading in time. We therefore expect that considering different, and possibly more realistic, matter sources for gravitational collapse would eliminate this effect.

We conclude by highlighting two important limitations of the model presented here. First, our analysis was restricted to the matter region, leaving aside the vacuum region exterior to it. Although the classical theory discussed in \cref{sec:2} can in principle also describe vacuum shells, it remains unclear whether their quantization can be carried out along the same lines as for the matter shells. While the matter region is undoubtedly central to understanding the physics of gravitational collapse, the vacuum region plays a significant role as well, and its inclusion in models of this kind is an important direction for future work.

The second limitation concerns the construction of the multi-shell model itself. Rather than stemming from a canonical quantization of the full marginally bound LTB model, it is assembled by treating the full system as a collection of independently quantized shells. This approach inherently overlooks potential inter-shell interactions of purely quantum origin, with no classical analogue, which could nonetheless prove decisive in shaping the global evolution of the multi-shell state and, in particular, the emergence of shell-crossing singularities. That said, given that a rigorous quantum treatment of the full theory is at present too involved to admit a thorough dynamical analysis, the multi-shell model stands as a useful and tractable framework for exploring the quantum evolution of multiple shells in a fully quantum gravitational setting. Moreover, if inter-shell interaction terms were to be identified in the Hamiltonian of the full quantum theory, they could be straightforwardly integrated into the multi-shell Hamiltonian and their effects studied in a controlled manner within this simplified setting.



\acknowledgments
The authors wish to thank Tomasz Pawłowski and Wojciech Kamiński for the valuable comments and discussions. The work was funded by the National Science Centre, Poland, as part of the OPUS 24 grant number 2022/47/B/ST2/02735.


\appendix 
\numberwithin{equation}{section}
\section{LTB conditions} \label{A}

In \cref{sec:2}, we introduced the LTB condition in \cref{LTBcondition} as an additional constraint of the theory. In the non-marginally bound case, it acts as a gauge condition for the radial diffeomorphism constraints, whereas in the marginally bound case it becomes an additional first-class constraint. This LTB condition effectively constrains the line element written in terms of the densitized triad in \cref{metricE},\footnote{The comoving gauge $N^x=0$ has already been imposed.}
\begin{equation} \label{metricEapp}
    ds^2 = -N^2 \, dt^2 + \frac{\:(E^\varphi)^2}{\abs{E^x}} dx ^2 + \abs{E^x} \,d\Omega^2 , 
\end{equation}
to take the LTB form given in \cref{metricLTB}, that is 
\begin{equation} \label{metricEapp2}
    ds^2 = -N^2 \, dt^2 + \frac{\left(\partial_x \sqrt{\abs{E^x}} \right)^2}{1 + \varepsilon(x)}  dx^2 + \abs{E^x} \,d\Omega^2 .
\end{equation}
There are, in fact, four different conditions that achieve this same result:
\begin{subequations} \label{C_LTB}
    \begin{align}
        C_{\mathrm{LTB}}^{(1)}&=E^{\varphi}-\frac{\partial_x E^x}{2 \sqrt{1+ \varepsilon (x)}}=0,\\
        C_{\mathrm{LTB}}^{(2)}&=E^{\varphi}-\frac{\partial_x \abs{E^x}}{2 \sqrt{1+ \varepsilon (x)}}=0,\\
        C_{\mathrm{LTB}}^{(3)}&=\abs{E^{\varphi}}-\frac{\partial_x E^x}{2 \sqrt{1+ \varepsilon (x)}}=0,\\
        C_{\mathrm{LTB}}^{(4)}&=\abs{E^{\varphi}}-\frac{\partial_x \abs{E^x}}{2 \sqrt{1+ \varepsilon (x)}}=0 \, .
    \end{align}
\end{subequations}
We chose the first condition $C_{\mathrm{LTB}}^{(1)}=0$ in the main part of the article, but any of the four conditions $C_{\mathrm{LTB}}^{(i)}=0$ could have been used equivalently. Each of them acts as a gauge condition for the radial diffeomorphism constraint in the non-marginally bound case, becomes a first-class constraint in the marginally bound case, and restrict the metric to the LTB form. In the marginally bound case, carrying out the same gauge-fixing procedure discussed at the end of \cref{sec:2.1} yields the reduced action
\begin{equation} \label{action_appendix}
    S=\frac{1}{2 \gamma G} \int dt \int dx \;  \partial_x \left( \alpha_{\mathrm{LTB}} \,\dot{K}_{\varphi} E^x +  \beta_{\mathrm{LTB}} \,\frac{K_{\varphi}^2 \sqrt{\abs{E^x}}}{\gamma} \right) ,
\end{equation}
where $\alpha_{\mathrm{LTB}}$ and $\beta_{\mathrm{LTB}}$ are coefficients encoding the different combinations of $\sgn(E^x)$ and $\sgn(E^\varphi)$ corresponding to the four choices of $C_{\mathrm{LTB}}^{(i)}$. Their values are listed in the following table:\\
\\
\setlength{\tabcolsep}{10pt} 
\renewcommand{\arraystretch}{1.5} 

\begin{table}[h]
\begin{tabular}{c|c|c|}
\cline{2-3}
                       & $\alpha_{\mathrm{LTB}}$ & $\beta_{\mathrm{LTB}} $ \\ \hline
\multicolumn{1}{|l|}{ $C^{(1)}_{\mathrm{LTB}}$ } & 1 & $\sgn(E^x) \sgn(E^\varphi)$ \\ 
\hline
\multicolumn{1}{|l|}{ $C^{(2)}_{\mathrm{LTB}}$ } & $\sgn(E^x)$ & $\sgn(E^\varphi)$ \\ 
\hline
\multicolumn{1}{|l|}{ $C^{(3)}_{\mathrm{LTB}}$ } & $\sgn(E^\varphi)$ & $\sgn(E^x)$ \\ 
\hline
\multicolumn{1}{|l|}{ $C^{(4)}_{\mathrm{LTB}}$ } & $\sgn(E^x) \sgn(E^\varphi)$ & 1 \\ 
\hline
\end{tabular}
\bigskip
\end{table}

\noindent
For $C^{(1)}_{\mathrm{LTB}}$, we naturally recover the result given in \cref{action_reduced3}. 

Each of these LTB conditions does not set a straightforward relation between the signs of $E^\varphi$ and $E^x$, rather, it relates the sign of $E^\varphi$ (or its absolute value) to the sign of $\partial_x E^x$ (or  $\partial_x \abs{E^x}$). For instance, choosing $C^{(1)}_{LTB}$ imposes $\sgn(E^\varphi)=\sgn(\partial_x E^x)$, that is, a positive (negative) $E^\varphi$ corresponds to a monotonically increasing (decreasing) $E^x$. Although $\sgn(E^x)$ is not fixed by this condition, we additionally require $E^x(x_{\mathrm c})=0$, since $\sqrt{\abs{E^x (x)}}$ represents the physical areal radius of the shell $x$. For this condition to be satisfied by an increasing $E^x$, we must have $\sgn(E^x)=1$; conversely, if $E^x$ is decreasing, $\sgn{E^x}=-1$.
All in all, this discussion leads to the condition $\sgn(E^\varphi)=\sgn(E^x)$ for $C_{LTB}^{(1)}$. The same argument can be applied to the three other choices of $C^{(i)}_{LTB}$. The results are summarized in the following table along with the new values of $\alpha_{LTB}$ and $\beta_{LTB}$.

\begin{table}[h]
\begin{tabular}{c|c|c|c|c|}
\cline{2-5}
                      & $\sgn{E^\varphi}$ &  $\sgn{E^x}$ & $\alpha_{\mathrm{LTB}}$ & $\beta_{\mathrm{LTB}} $ \\ \hline
\multicolumn{1}{|l|}{$C^{(1)}_{\mathrm{LTB}}$}
                      & +1 / -1 &  +1 / -1 & 1 &  $1$ \\ \hline
\multicolumn{1}{|l|}{ $C^{(2)}_{\mathrm{LTB}}$ }  
                      & +1 & $\pm$ 1 & $\sgn(E^x)$ & 1 \\ \hline
\multicolumn{1}{|l|}{ $C^{(3)}_{\mathrm{LTB}}$ } 
                      & $\pm$ 1 & +1 & $\sgn(E^\varphi)$ & 1 \\ \hline
\multicolumn{1}{|l|}{ $C^{(4)}_{\mathrm{LTB}}$ }  
                      &  +1 / -1 & $\pm$ 1 / $\pm$ 1 & $\sgn(E^x)\sgn(E^\varphi)$ & 1 \\ \hline
\end{tabular}
\bigskip
\end{table}

Interestingly, $\beta_{LTB}=1$ for all choices of $C_{LTB}^{(i)}$, namely the Hamiltonian keeps always the same sign.
A second comment can be made regarding the degeneracy associated with the choice of the LTB constraint. Given that the metric does not depend on the signs of $E^\varphi$ and $E^x$, the first three choices of $C_{LTB}^{(i)}$ give a 2-fold degeneracy whereas the last one produces 4 equivalent sectors of the theory (maximal degeneracy).

\newpage
\section{Self-adjointness of the WDW and LQG Hamiltonians} \label{B}

\subsection{Wheeler-DeWitt Hamiltonian}

In order for the Wheeler–DeWitt states to evolve unitarily in time, the evolution must be generated by a self-adjoint operator. We now show that, although $\underline{\hat{H}}$ is not self-adjoint, it admits a one-parameter family of self-adjoint extensions. To begin with, $\underline{\hat{H}}$ is manifestly a symmetric operator. We can therefore define its deficiency subspaces $\mathcal K_\pm$ as the spaces spanned by the normalizable solutions of
\begin{equation} \label{deficiencyeq}
    \underline{\hat{H}} {}^{\dagger} {\mathrm e}_{\pm i} = \pm i \, {\mathrm e}_{\pm i} \, .
\end{equation}
Let $n_\pm$ be the dimension of $\mathcal K_\pm$. As discussed in~\cite{reed1975ii}, $\underline{\hat{H}}$ is self-adjoint if and only if $n_+=n_-=0$; it admits self-adjoint extensions if and only if $n_+=n_-$; otherwise, it is neither self-adjoint nor extendable to a self-adjoint operator. In the present case, the only two normalizable solutions to \cref{deficiencyeq} are
\begin{equation}
    {\mathrm e}_{\pm i}=\frac{N}{\abs{v}^{\frac{1}{4}}} e^{-(1 \mp i)\sqrt{\frac{2 \gamma  \sqrt{\Delta}}{3 \hbar}} \sqrt{\abs{v}}} \, ,
\end{equation}
from which it follows that $n_+=n_-=1$. Consequently, $\underline{\hat{H}}$ admits (a one-parameter family of) self-adjoint extensions.

Let $\hat{S}$ be a self-adjoint extension of $\underline{\hat{H}}$. Since $\hat{S}$ is self-adjoint, it must satisfy 
\begin{equation}
   \big(\varphi, \hat{S} \varphi\big) - \big(\hat{S} \varphi , \varphi\big) =0 
\end{equation} for all $\varphi \in D(\hat{S})$. An explicit evaluation of this condition shows that there is actually a one-parameter family of self-adjoint extensions $\underline{\hat{H}}_a$ of $\underline{\hat{H}}$, labelled by a parameter $a\in \mathbb{R} \,\cup \,\{\infty\}$. Their domains are given by
\begin{equation}
        D(\underline{\hat{H}}_{a})=\Big\{\varphi(v) \in \mathcal{S}(\mathbb{R}) : \;\varphi(v)=\varphi(-v),\;\; \lim_{\;\,v\rightarrow 0^+}\left[v^\frac{1}{2} \partial_v \left(v^\frac{1}{4} \varphi(v)\right) - a \, v^\frac{1}{4}\varphi(v)\right]=0 \Big\} 
\end{equation}
for $a\in\mathbb{R}$, and 
\begin{equation}
        D(\underline{\hat{H}}_{\infty})=\Big\{\varphi(v) \in \mathcal{S}(\mathbb{R}) : \;\varphi(v)=\varphi(-v),\;\; \lim_{\;\,v\rightarrow 0^+}\left[ v^\frac{1}{4}\varphi(v)\right]=0 \Big\} 
\end{equation}
for $a=\infty$. Despite its seemingly unusual form, the boundary condition defining $D(\underline{\hat{H}}_{a})$ is entirely analogous to the one encountered, for instance, in the study of self-adjoint extensions of the Hamiltonian of a free particle on the half line (see Example 2 of Chapter X in~\cite{reed1975ii}). As in that case, a basis $\underline{\mathrm e}_{a,k}$ for $D(\underline{\hat{H}}_{a})$ is provided by linear combinations of the generalized eigenfunctions $\underline{\mathrm{e}}_k$ of $\underline{\hat{H}}$,
\begin{equation}
    \underline{\mathrm e}_{a,k}(v) \coloneq \underline{\mathrm{e}}_{-k}(v) + \alpha (a,k)\,\underline{\mathrm{e}}_k (v)= \frac{N}{\abs{v}^{\frac{1}{4}}} \Big(e^{-i k \sqrt{\abs{v}}} +\alpha (a,k)\, e^{i k \sqrt{\abs{v}}} \Big) ,
\end{equation}
where the coefficient $\alpha(a,k)$ is fixed by the boundary condition and is given by $\alpha=(ik+2a)/(ik-2a)$. The state $\underline{\mathrm e}_{a,k}$ can be interpreted as an ingoing wave that is reflected, i.e. ``bounces'', at the origin, with the reflected component acquiring a phase $\alpha$. Since $\alpha$ is a pure phase, $\abs{\alpha}=1$, we may write $\alpha=e^{i\beta}$, which implies $a=-(k/2) \tan\left(\beta/2\right)$ and allows us to rewrite the generalized eigenfunctions of $\underline{\hat{H}}_a$ as 
\begin{equation} \label{eakWDW}
    \underline{\mathrm e}_{a, \, k} (v) = \frac{1}{\sqrt{2 \pi}\abs{v}^{\frac{1}{4}}}  \cos\left(k \sqrt{\abs{v}} + \frac{\beta(a,k)}{2} \right).
\end{equation}
The normalization constant has been fixed to $N=1/\sqrt{2 \pi}$ so that $\underline{\mathrm e}_{a, \, k}$ is Dirac-delta normalized for the special cases $a=0$ and $a=\infty$. An arbitrary state $\varphi \in D(\underline{\hat{H}}_a)$ can then be written as
\begin{equation}
    \varphi(v)=\int_{0}^{+\infty} dk \, \,\widetilde{\varphi} (k) \, \underline{\mathrm e}_{a, \, k} 
\end{equation}
for some spectral profile $\widetilde{\varphi} (k) $.

\subsection{LQG Hamiltonian}

We now prove that the LQG Hamiltonian $\hat{H}$, unlike the WDW one $\underline{\hat{H}}$, is essentially self-adjoint.
To do so, we will make use of lemma 2.16, point (ii), in \cite{teschl2000jacobi}.
The starting point is to construct the Jacobi equation for a given $z \in \mathbb{C}$
\begin{equation} \label{Jacobi}
    \xi(v+4) = \frac{z-f_0(v)}{f_+(v)} \xi(v)- \frac{f_-(v)}{f_+(v)} \xi(v-4) \, ,
\end{equation}
defined in $L_\epsilon \owns v \ge 4+\epsilon$ and equipped with two initial conditions $\xi(\epsilon)$ and $\xi(\epsilon+4)$.
Since solutions rely on two initial conditions, one can construct two linearly independent solutions $A(v)$ and $B(v)$ with
\begin{align*}
    &A(\epsilon)=0 \quad\quad\quad B(\epsilon)=1\\
    &A(\epsilon+4)=1 \quad\; B(\epsilon+4)=0\,
\end{align*}
which then span a bidimensional space.
As a side remark, the loop eigenfunctions satisfy \eqref{ek_recursive}  which is identical to  \eqref{Jacobi} but for the additional property
\begin{equation}
    e_k(\epsilon+4)=-\frac{\omega_k^2+f_0(\epsilon)}{f_+(\epsilon)} e_k(\epsilon) \, ,
\end{equation}
obtained by setting $v=\epsilon$ in \eqref{ek_recursive} and by noting that $f_-(\epsilon)=0$.
Therefore, $e_k(v)$ are particular solutions of \eqref{Jacobi}.

By lemma 2.16 in \cite{teschl2000jacobi}, $\hat{H}$\footnote{In \cite{teschl2000jacobi}, the lemma is formulated for an operator $H_{max, +}$ which, by its definition (1.90), is analogous to the LQG Hamiltonian.} is self-adjoint if and only if there exists a solution to \eqref{Jacobi} --- for any $z \in \mathbb{C}$ --- which is not square summable near infinity. 
As a remark, given that $f_+(v\ge \epsilon +4) \neq 0$, none of the points $\xi(v)$ is singular. If one shows that at least one of the solutions to the Jacobi equation has infinite norm, it necessarily implies that it is not square summable near $+\infty$. Therefore, to show that $\hat{H}$ is self adjoint, it suffices to prove that $\norm{\xi}=\infty$ for a given $\xi$ solution to \eqref{Jacobi}. 
To this end, pick two linearly independent solutions to the Jacobi equation, $\psi(v)$ and $\phi(v)$, and construct the quantity
\begin{equation} \label{wronskian}
    W(v) \coloneqq f_+(v)  \left[\psi(v+4) \phi(v) - \psi(v) \phi(v+4)   \right] \, .
\end{equation}
It is easy to show that $W(v)=W(v+4n) \neq 0$ and it is therefore a complex constant.
We then have
\begin{equation}
    \begin{split}
        \abs{\frac{W}{f_+(v)}} &\le  \abs{\psi(v+4)} \, \abs{\phi(v)} + \abs{\psi(v)} \, \abs{\phi(v+4)}\\
        &\le \frac{1}{2} \left( \abs{\psi(v+4)}^2 + \abs{\phi(v)}^2 + \abs{\psi(v)}^2 +  \abs{\phi(v+4)} ^2\right) \, .
    \end{split}
\end{equation}
Summing over $v$ gives
\begin{equation}
        \abs{W} \sum_{v > 0}^{+\infty} \abs{\frac{1}{f_+(v)}} \le \norm{\psi}^2 + \norm{\phi}^2 \, .
\end{equation}
In the large $v$ limit, $f_+(v) \approx v$, which implies $\sum_{v > 0}^{+\infty} \abs{\frac{1}{f_+(v)}} =+\infty$. As the RHS of the last inequality must then be $+\infty$ as well, at least one of the two solutions has infinite norm. \qed

\section{Large \texorpdfstring{$v$}{v} behavior and normalization of the loop eigenfunctions} \label{C}

In the large $v$ regime, where the discreteness of the underlying geometry becomes negligible, the LQG Hamiltonian $\hat{H}$ reduces to its WDW counterpart $\underline{\hat{H}}$~\cite{Ashtekar_2006_2}. It is therefore natural to expect that, in the same regime, the loop eigenfunctions $\mathrm e_{k}(v)$ reduce to their WDW counterparts $\underline{\mathrm e}_{k}(v)$. That this is indeed the case was demonstrated in~\cite{Kaminski_2010} in the context of a homogeneous universe with a scalar field. The same analysis can be adapted to the present setting.

Let us reformulate the problem in a matrix formalism, in which it takes a particularly simple form. Define
\begin{equation}
    \vec{\mathrm e}_{k}(v)\coloneq
    \begin{pmatrix} 
        \mathrm e_{k} (v) \\  \mathrm e_{k} (v-4) 
    \end{pmatrix} 
    \quad\quad \text{and} \quad\quad
    \mathbb{A}_{k} (v)\coloneq
    \begin{pmatrix}
            -\frac{ \omega_k^2 + f_0(v)}{f_+(v)} & -\frac{f_-(v)}{f_+(v)} \\
        1 & 0
    \end{pmatrix}  ,
\end{equation}
so that \cref{ek_recursive} can be rewritten compactly as 
\begin{equation}\label{matrix_recursive_eq}
    \vec{\mathrm e}_{k}(v+4) = \mathbb{A}_{k}(v)\, \vec{\mathrm e}_{k}(v)\,.
\end{equation}
Next, introduce the matrix $\mathbb{B}_{k}$ built from the WDW eigenfunctions,
\begin{equation}
    \mathbb{B}_{k}(v)\coloneq
    \begin{pmatrix}
        \,\underline{\mathrm e}_{k}(v) & \underline{\mathrm e}_{k}^*(v)\, \\
        \underline{\mathrm e}_{k}(v-4) & \underline{\mathrm e}_{k}^*(v-4) 
    \end{pmatrix}  ,
\end{equation}
and define a new vector $\vec{\chi}_{k}(v)$ through
\begin{equation} \label{chi_def}
    \vec{\mathrm e}_{k}(v)=\mathbb{B}_{k}(v)\,\vec{\chi}_{k}(v) \,.
\end{equation}
It can be verified analytically that the matrix $\mathbb{B}_{k}(v)$ is always invertible for $v$ bigger than a given point $v_k$ which is $k$-dependent. Therefore, in the large $v$ limit, $\vec{\chi}_{k}(v)$ is always well defined. Since the functions $\mathrm e_{k}(v)$ are real, the vector $\vec{\chi}_{k}(v) $ must take the form
\begin{equation}
     \vec{\chi}_{k}(v)= r(k, v)
    \begin{pmatrix} 
        e^{i \alpha(k, v)} \\   e^{-i \alpha(k, v)}
    \end{pmatrix} 
\end{equation}
for some real-valued functions $r(k,v)$ and $\alpha(k,v)$. If $r(k,v)=r_k$ and $\alpha(k,v)=\alpha_k$ are independent of $v$, then $\mathrm e_{k}(v)$ reduces to a simple linear combination of $\underline{\mathrm e_{k}} (v)$ and $\underline{\mathrm e^*_{k}}(v)$. In general, since $\mathrm e_{k}(v)$ and $\underline{\mathrm e_{k}}(v)$ are not simply related, the functions $r(k,v)$ and $\alpha(k,v)$ depends non-trivially on $v$ and parametrize this difference. However, if the loop eigenfunctions $\mathrm e_{k}(v)$ approach a linear combination of $\underline{\mathrm e_{k}} (v)$ and $\underline{\mathrm e^*_{k}}(v)$ in the regime $v\gg1$, then the limit for $v\to\infty$ of $\vec{\chi}_{k}(v) $ must exist and be equal to a constant vector $\vec{\chi}_{k}$.

This is indeed the case, and it becomes particularly transparent within the matrix formalism. Using \cref{matrix_recursive_eq,chi_def}, we obtain
\begin{equation} \label{M_def}
    \vec{\chi}_{k}(v+4) = \mathbb{B}_{k}^{-1}(v+4) \mathbb{A}_{k}(v) \mathbb{B}_{k}(v)\, \vec{\chi}_{k}(v) \eqqcolon \mathbb{M}_{k}(v) \,\vec{\chi}_{k}(v) \, .
\end{equation}
As argued in~\cite{Kaminski_2010}, a sufficient condition for
\begin{equation}
    \lim_{v\to\infty} \vec{\chi}_{k}(v) = \vec{\chi}_{k}
\end{equation}
to exist is that
\begin{equation} \label{M_asymptotic}
    \mathbb{M}_{k}(v) = \mathds{1} + \mathbbm{o} (v^{-1})
\end{equation}
as $v\gg 1$. A direct inspection of $\mathbb{M}_{k}(v)$, using the explicit expressions for $f_{\pm,0}(v)$ and $\underline{\mathrm e_{k}}(v)$ given in \cref{sec:3}, shows that
\begin{equation}
    \mathbb{M}_{k}(v) = \mathds{1} + \mathbb{O} (v^{-3/2})
\end{equation}
as $v\gg 1$, thereby proving the asymptotic relation between loop and WDW eigenfunctions. Concretely, one finds
\begin{equation} \label{large_v_limit}
    \mathrm e_k (v) = r(k) \Big( e^{i\alpha(k)} \underline{\mathrm e}_{k}(v) + e^{-i\alpha(k)} \underline{\mathrm e}^*_{k}(v)\Big) + O \Big(\abs{\underline{\mathrm e}_{k}(v)} v^{-1/2}\Big)
\end{equation}
as $v\gg 1$. Moreover, it can be shown (see Appendix A2 in~\cite{Kaminski_2010}) that the normalization condition $\braket{\mathrm e_k}{\mathrm e_{k'}}=\delta(k-k')$ further constrains $r(k) $ to be $r(k) =2$.

This is a fundamental result, as it allows a straightforward normalization of the numerically computed eigenfunctions $\mathrm e_k (v)$. As discussed in \cref{sec:3}, the $\mathrm e_k (v)$ can be chosen to be Dirac-delta normalized by appropriately fixing the value of $\mathrm e_k(\epsilon)\in\mathbb{R}^+$. Determining this value analytically, however, is rather involved. Instead, we numerically construct a non-normalized eigenfunction $\tilde{\mathrm e}_k (v)$ by solving the recurrence relation in \cref{ek_recursive} with the initial condition $\mathrm e_k(\epsilon)=1$. In the large-$v$ regime, $\tilde{\mathrm e}_k (v)$ approaches an expression of the form given in \cref{large_v_limit}, with a scaling factor $\tilde{r}(k)$ that can be extracted numerically. A properly normalized eigenfunction $\mathrm e_k (v)$ is then obtained through the simple rescaling $\mathrm e_k (v) = 2\!\: \tilde{\mathrm e}_k (v) /\tilde{r}(k)$.

Given the order of the subleading terms in \cref{M_asymptotic,large_v_limit}, the numerical asymptotic analysis described above, although straightforward to implement, exhibits a relatively slow convergence rate. This can be improved by going beyond the leading-order approximation and retaining sufficiently many subleading terms to achieve the desired rate of convergence. Concretely, following~\cite{Paw_owski_2014}, this is accomplished by introducing the modified WDW functions $\underline{\mathrm e}^{(q)}_k (v)$, valid for $v \gg 1$, defined as
\begin{equation}
    \underline{\mathrm e}^{(q)}_k (v) \coloneq \frac{1}{\sqrt{2 \pi}} \left( \sum_{j=0}^{J_1} \frac{B_j}{v^{\frac{1}{4}+j}} \right) \exp\left(i k \sum_{j=0}^{J_2} A_j \,v^{\frac{1}{2}-j} \right) ,
\end{equation}
where the $J_1$ coefficients $A_j$ and the $J_2$ coefficients $B_j$ ($A_0=B_0=1$) are determined by requiring that the matrix $\mathbb{M}^{(q)}_{k}(v)$, constructed from the $\mathbb{B}^{(q)}_{k}(v)$ matrix of the modified basis $\underline{\mathrm e}^{(q)}_k (v)$, satisfy
\begin{equation}
    \mathbb{M}^{(q)}_{k}(v) = \mathds{1} + \mathbb{O} (v^{-q})
\end{equation}
as $v \gg 1$. For $q=4$, one obtains
\begin{equation}
    \underline{\mathrm e}^{(4)}_k = \frac{1}{\sqrt{2 \pi}} \left(\frac{1}{v^{1/4}} + \frac{B_1}{v^{5/4}} +\frac{B_2}{v^{9/4}} \right) \exp\left[i k \left( v^{1/2} + \frac{A_1}{v^{1/2}}  + \frac{A_2}{v^{3/2}}  + \frac{A_3}{v^{5/2}} \right) \right]  ,
\end{equation}
with
\begin{equation} \label{parameters_AB}
    \begin{split}
    A_1 &=-\frac{1}{6}k^2, \\
    A_2 &=\frac{1}{6}-\frac{1}{40}k^4, \\
    A_3 &=\frac{37}{120} k^2-\frac{1}{112}k^6,\\
    B_1&=\frac{1}{4}k^2,\\
    B_2&=\frac{5}{32} k^4\, .
    \end{split}
\end{equation}
The asymptotic expression in \cref{large_v_limit} then improves to
\begin{equation} \label{large_v_limit_2}
    \mathrm e_k (v) = 2 \Big( e^{i\alpha(k)} \underline{\mathrm e}^{(4)}_{k}(v) + e^{-i\alpha(k)} \underline{\mathrm e}^{(4)}_{k}{}^*(v)\Big) + O \Big(\abs{\underline{\mathrm e}^{(4)}_{k}(v)} v^{-3}\Big)
\end{equation}
for $v \gg 1$.

\section{Semiclassical effective theory} 
\label{D}

In the following appendix, we briefly summarize the relevant results of the LQG semiclassical theory for the purpose of comparison with the full quantum dynamics.

The underlying idea of this construction is to use approximation methods to analytically estimate the dynamics of the expectation values of the elementary quantum operators on a semiclassical state, and to subsequently interpret the result as a quantum-corrected effective evolution unfolding in the classical phase space. The rigorous derivation of this procedure is discussed in~\cite{Ashtekar:2011ni}. In many situations, including the one considered here, this effective dynamics in the classical phase space can be seen as the evolution generated by a semiclassical Hamiltonian $H_{sc}$ obtained from the classical Hamiltonian $H$ via the heuristic substitution (see also \cref{b_to_sin})
\begin{equation}
    b \rightarrow \sin(b) \, ,
\end{equation}
commonly referred to as the ``polymerization'' of $b$. This yields
\begin{equation} \label{H_sc}
    H_{sc}=-\frac{3}{4} \frac{\hbar}{\gamma \sqrt{\Delta}} \sin(b)^2 \abs{v} \, .
\end{equation}

The effective EOMs for a given shell $x$ are derived by following the same procedure as in the classical theory, namely by computing the Poisson brackets
\begin{align}
    \dot{b} &= \{b, H_{sc} \}= -\frac{3}{2 \gamma \sqrt{\Delta}} \sgn(v) \sin^2(b),\\
    \dot{v} &= \{v, H_{sc} \}= \frac{3}{\gamma \sqrt{\Delta}} \abs{v} \sin(b) \cos(b) \, .
\end{align}
Throughout the following, we restrict to positive $v$, without any loss of generality. Introducing the semiclassical global mass-energy density $\rho_{sc}= -H_{sc}/V$, the equation of motion for $v$ takes the form
\begin{equation} \label{EOM_v_sc}
    \left(\frac{\dot{v}}{v}\right)^2 = 24 \pi G \rho_{sc}\left(1- \frac{\rho_{sc}}{\rho_c}\right)  .
\end{equation}
Since the total mass-energy enclosed within a shell is conserved also in the semiclassical theory, we have $\rho_{sc}\propto v^{-1}$, and \cref{EOM_v_sc} admits the analytical solution
\begin{equation} \label{v_sc}
    \frac{v(t)}{v_0}= 1 -  \sqrt{24 \pi G \rho_0 \left(1- \frac{\rho_0}{\rho_c}\right)} \:t + 6 \pi G \rho_0 \,t^2   .
\end{equation}
This solution describes an initially collapsing shell that reaches a non-zero minimum volume at 
\begin{equation}
    T_b=\sqrt{\frac{1- \rho_0/\rho_c}{6 \pi G \,\rho_0}}\, ,
\end{equation}
thereby avoiding the classical singularity, before bouncing back and subsequently expanding.

The extent to which this semiclassical dynamics faithfully captures the evolution of sufficiently peaked wave packets in the loop quantum theory is assessed in \cref{sec:3.2}.

\newpage
\bibliography{main}

\end{document}